\RequirePackage{fix-cm}
\documentclass[smallextended]{svjour3} 
\smartqed  
\usepackage{cite}
\usepackage[table]{xcolor}
\usepackage{amsmath,amssymb,amsfonts}
\usepackage{algorithmic}
\usepackage{graphicx}
\usepackage{textcomp}
\usepackage{booktabs}
\usepackage{amsmath, balance}
\usepackage{scalerel}
\usepackage{tikz}
\usetikzlibrary{svg.path}
\usepackage{arydshln}
\usepackage{caption}
\usepackage{academicons}
\usepackage{array}  
\usepackage{lipsum}  
\usepackage{adjustbox}
\usepackage{mdframed}
\usepackage{multirow}
\usepackage{url}
\usepackage[most]{tcolorbox}
\usepackage{enumitem}
\usepackage{ragged2e}
\usepackage{lmodern}
\usepackage[T1]{fontenc}
\usepackage{xcolor}
\usepackage{listings}
\usepackage{tabularx}
\usepackage{makecell}
\usepackage{subfig}
\usepackage{arydshln}

\definecolor{Gr}{rgb}{0.0, 0.5, 0.0}
\definecolor{light-gray}{gray}{0.95}
\newcommand{\code}[1]{\colorbox{light-gray}{\texttt{#1}}}

\newcommand{\specialcell}[2][c]{%
  \begin{tabular}[#1]{@{}c@{}}#2\end{tabular}}

\lstdefinestyle{diffstyle}{
  backgroundcolor=\color{gray!7},
  basicstyle=\ttfamily\scriptsize,
  frame=single,
  rulecolor=\color{gray!25},
  breaklines=true,
  showstringspaces=false
}

\tcbset{samplebox/.style={
  colback=gray!3, colframe=black!35, boxrule=0.6pt,
  arc=2mm, left=6pt, right=6pt, top=6pt, bottom=6pt,
  enhanced, before skip=10pt, after skip=10pt
}}

\newlist{compactenum}{enumerate}{1}
\setlist[compactenum]{nosep, leftmargin=*, label=\arabic*., itemsep=1pt, topsep=2pt}
\newcolumntype{L}{>{\centering\scriptsize\bfseries}p{0.18\textwidth}}
\newcolumntype{Y}{>{\raggedright\arraybackslash}X}
    
\begin{document}

\title{Analysis of AdvFusion: Adapter-based Multilingual Learning for Code Large Language Models}

\author{Amirreza Esmaeili  \and
        Fahd Seddik             \and
        Yongyi Ji               \and
        Fatemeh Fard            \and
        Fuxiang Chen
}


\date{Received: date / Accepted: date}

\institute{
            Amirreza Esmaeili \at
              University of British Columbia \\
              3333 University Way, Kelowna, BC V1V 1V7, Canada \\
              \email{a.esmaeili@ubc.ca}         
           \and
            Fahd Seddik \at
              University of British Columbia \\
              3333 University Way, Kelowna, BC V1V 1V7, Canada \\
              \email{fahd.seddik@ubc.ca}           
           \and
            Yongyi Ji  \at
              University of Leicester \\
              University Rd, Leicester LE1 7RH, United Kingdom \\
              \email{yj171@leicester.ac.uk}           
           \and
            Fatemeh Fard \at
              University of British Columbia \\
              3333 University Way, Kelowna, BC V1V 1V7, Canada \\
              \email{fatemeh.fard@ubc.ca}           
           \and
           Fuxiang Chen \at
              University of Leicester \\
              University Rd, Leicester LE1 7RH, United Kingdom \\           
              \email{fuxiang.chen@leicester.ac.uk}           
}

\maketitle

\begin{abstract}
Programming languages can benefit from one another by utilizing a language model for software engineering tasks. Full fine-tuning and Parameter Efficient Fine-Tuning (PEFT) of Code Language Models (Code-LMs) has been explored for multilingual knowledge transfer. 
AdapterFusion is a PEFT architecture that aims to enhance task performance by leveraging information from multiple programming languages, but primarily focuses on the target programming language.

In our previous work, we proposed \textbf{AdvFusion}, a novel PEFT-based approach that effectively learns from other programming languages before adapting to the target task. 
Though previous experiments showed that AdvFusion outperformed AdapterFusion and LoRA, it was applied on pre-trained Code-LMs and was limited to only two tasks, code summarization and method name prediction. 
In this study, we expanded our work and investigated AdvFusion on Code Large Language Models (Code-LLMs), considering three new tasks: code generation, code translation, and commit message generation. 
We observed that different Code-LLMs/tasks exhibit different characteristics. In code generation, AdvFusion outperformed AdapterFusion but not other PEFT methods (LoRA, Compacter, and TaskAdapter). In commit message generation, AdapterFusion performed better than AdvFusion, and contrary to code generation, we found that the other PEFT methods do not have better performance. In code translation, AdvFusion performed worse than AdapterFusion overall, with the performance gap marginally widening as the model size increases. However, consistent with code generation, other PEFT methods showed better performance.
\end{abstract}

\section{Introduction}\label{sec:introduction}

Parameter Efficient Fine-Tuning (PEFT) approaches, computationally efficient alternatives to full fine-tuning, are well-known techniques that fine-tune a language model on a small set of weights~\cite{houlsby2019parameter, wang2023oneAdapter, zhuo2024astraios}, and they have been widely used in Software Engineering (SE) studies \cite{zhuo2024astraios,weyssow2023exploring,liu2023empirical, wang2023oneAdapter}. Several studies, both in SE and Natural Language Processing (NLP), utilized various PEFT architectures to adapt a language model to downstream tasks~\cite{goel2022cross, pfeiffer2020mad,pfeiffer2020adapterfusion,hu2021lora,rathnayake2022adapter,zhuo2024astraios,weyssow2023exploring,liu2023empirical,honarvar2025turbulence,machavcek2025impact}.

Among PEFT architectures, AdapterFusion~\cite{pfeiffer2020adapterfusion}, based on Adapter modules inserted between the Transformer layers~ \cite{houlsby2019parameter}, is trained with the purpose of enhancing the performance of a target task in a specific language by leveraging similar latent information from other languages. This knowledge transfer among languages is particularly important for low-resource languages, those for which the amount of training data is limited~\cite{ahmed2022learning}.
Though AdapterFusion is inherently developed to learn from different languages, in our previous work~\cite{saberi2025advfusion}, our experiments revealed that with this architecture, the models are still learning mainly from the same programming language of the target task, rather than from other programming languages. We therefore proposed \textbf{Adversarial Fusion Adapter (AdvFusion)}, a PEFT architecture that enforces AdapterFusion to first learn from other programming languages before attending to the programming language of the target task. Thus, AdvFusion enhanced the knowledge transfer among programming languages~\cite{saberi2025advfusion}.

AdvFusion was previously evaluated on two tasks, code summarization and method name prediction, and six programming languages, Java, Python, PHP, JavaScript, Go, and Ruby \cite{saberi2025advfusion}. Our experiments showed that AdvFusion significantly enhances the performance of multilingual PEFT approaches.
In code summarization, we observed a $10\%$ improvement in BLEU across various models. Additionally, AdvFusion boosted method name prediction performance, achieving up to a $9\%$ increase in F1-score compared to AdapterFusion. Furthermore, AdvFusion outperformed LoRA, a widely recognized PEFT technique, by up to $12\%$ in BLEU and $32\%$ in F-1 for code summarization and method name prediction, respectively~\cite{saberi2025advfusion}.
We further observed that AdvFusion enhanced the performance of low-resource programming languages in some cases.

\noindent\textbf{Extensions from Prior Publication.}
Building on our previous work, in this invited extension of the AdvFusion paper~\cite{saberi2025advfusion}, we introduced new tasks and new foundational models to compare to explore whether AdvFusion perform effectively on Code-LLMs for low-resource programming languages.
For the new tasks, we studied the three tasks of code generation, code translation, and commit message generation, which reflect the challenges developers encountered in real-world work. The three tasks are applied on CodeLlama \cite{code-llama-roziere2023code}, DeepSeek-Coder \cite{zhu2024deepseek}, Qwen2.5-Coder \cite{hui2024qwen2} and their variants, given their high popularity within the Software Engineering community.
To evaluate AdvFusion, we compared it with several popular PEFT methods, including AdapterFusion \cite{pfeiffer2020adapterfusion}, LoRA \cite{hu2021lora}, Compacter \cite{karimi2021compacter}, and TaskAdapter \cite{houlsby2019parameter}.
We applied different popular and widely used datasets for different tasks. The selected datasets include multilingual and low-resource programming languages. Specifically, we used the xCodeEval dataset \cite{khan-etal-2024-xcodeeval} for code generation, the CommitPackFT dataset \cite{muennighoff2023octopack} for commit message generation, and the CodeTransOcean dataset \cite{yan2023codetransocean} for code translation.

We observed that different tasks exhibit different results. In code generation, AdvFusion consistently performed better than AdapterFusion across all Code-LLMs. The other PEFT methods performed better than AdvFusion in multiple cases; for example, TaskAdapter achieved the best performance in Deepseek-Coder, Qwen2.5-Coder 1.5B and CodeLlama. Contrary to code generation, AdapterFusion frequently outperformed AdvFusion and other PEFT methods in commit message generation. In code translation, Advfusion did not perform as well as AdapterFusion, and the performance gap between Advfusion and AdapterFusion marginally widened as the model size increased. Similar to the result of the other PEFT methods in code generation, LoRA outperformed AdvFusion and AdapterFusion in code translation.
Nonetheless, our experiments showed that in some situations, AdvFusion does indeed capture more knowledge learnt from other programming languages than AdapterFusion. 
We also observed that adapters with simple architecture, such as TaskAdapter, can achieve better performance than adapters with a more complicated architecture, such as LoRA, when used on Code-LLMs in code generation and commit message generation. This result and insight thus call for researchers to propose more efficient adapter architecture designs for Code-LLMs, i.e., adapters that are more complex and perform better in other domains do not necessarily work similarly on code-related tasks.

Our contributions in this extended paper are as follows. 
\begin{itemize}
    
    \item {Extending the study of AdvFusion for Code-LLMs and assessing their capabilities in transferring knowledge among programming languages, with a focus on low-resource ones. }
    \item {Empirically study the capabilities of AdvFusion compared to other PEFT methods for three tasks, code generation, code translation, and commit message generation, which represent NL-code, code-code and {code-NL}. }
\end{itemize}

The paper is organized as follows. For the purpose of completeness of this paper, we kept some of the important parts of the earlier AdvFusion work (Sections~\ref{section:background}, \ref{section:approach}, and \ref{section:advfusion-setup-on-codelms}). In Section \ref{section:background}, we provide an overview of the necessary background information.
We introduce our novel PEFT architecture, AdvFusion, in Section \ref{section:approach}. 
We provide the experimental setup and results for Code-LMs in Section \ref{section:advfusion-setup-on-codelms}, which is mainly adopted from our previous work. 
The new experiments in this extension work are presented in Section 5 onwards. 
We provide the research questions and experimental setup for Code-LLMs in Section \ref{section:advfusion-setup-on-codellms}. Section \ref{section:advfusion-results-on-codellms} describes the results, and they are further discussed in Section \ref{section:discussion}.
Sections \ref{section:related-work} and \ref{section:threats} are dedicated to the related works and threats to validity. Finally, we conclude the paper in Section \ref{section:conclusion}. 

\begin{figure}
\centering
    \scalebox{.5}{
    \includegraphics{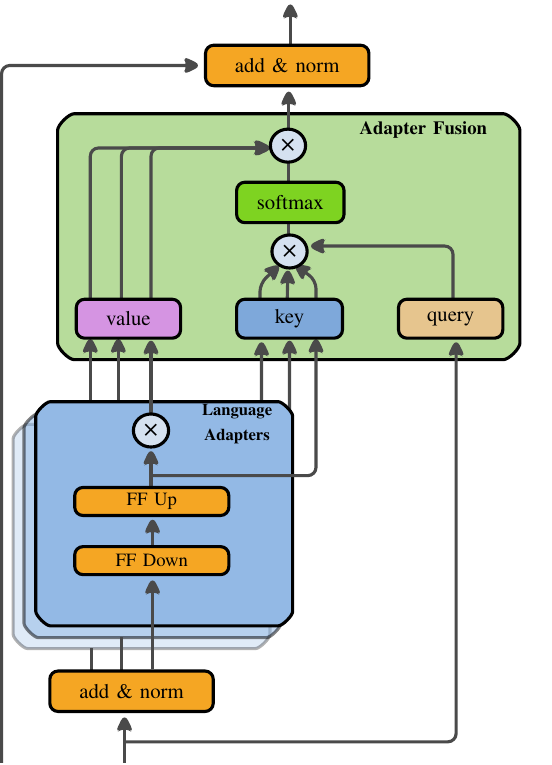}
    }
\caption{Internal structure of AdapterFusion. }

\label{fig:fusion-internal}
\end{figure}

\section{Background}
\label{section:background}

\subsection{Adapters} 
\label{subsection:adapters}

We used various adapter types in our approach: task adapters, language adapters, and AdapterFusions. Adapters are lightweight modules added to a language model's internal structure, providing an efficient alternative to traditional fine-tuning for new tasks and preventing catastrophic forgetting. Adapters require less computational time and resources than fine-tuning.

Let $\Theta$ represent all weights of a pre-trained model. When an adapter $i$ is added, a new set of weights $\theta_i$ is created. During training, $\Theta$ remains frozen, and only $\theta_i$ is trained for the downstream task.

\subsubsection{Task Adapters}

The aim of a task adapter is to learn a task-specific functionality by training its weights on a target task dataset \cite{pfeiffer2020mad}. Task adapters consist of a  simple down- and up-projection combined with residual connections. Task adapter $TA_l$ at layer $l$ consists of a down-projection $D \in R^{h\times d} $  where $h$ is the hidden size of the Transformer, $d$ is the dimension of the adapter, and $r_l$ represents the residual connections at layer $l$. The down-sampled representations are then fed to a ReLU activation followed by an up-projection transformation $U \in R^{d\times h} $ at each layer. This is shown in Equation \ref{eq:task-adapter}:

\begin{equation}
TaskAdapter_l(h_l,r_l)=U_l(ReLU(D_l(h_l)))+r_l
\label{eq:task-adapter}
\end{equation}

\subsubsection{Language Adapters}
Language adapters learn language-specific features by training their weights on an abstract objective function such as MLM \cite{pfeiffer2020mad}. The language adapter $LA_l$ at layer $l$ has the same architecture as a task adapter. The internal structure of a language adapter consists of a down-projection $D \in R^{h\times d} $ with a ReLU activation, followed by an up-projection $U \in R^{d\times h} $, as shown in Equation \ref{eq:language-adapter}:

\begin{equation}
LanguageAdapter_l(h_l,r_l)=U_l(ReLU(D_l(h_l)))+r_l
\label{eq:language-adapter}
\end{equation}

where $h_l$ and $r_l$ are defined similarly to task adapters. 
Language adapters differ from task adapters in that they are trained on unlabeled data using Masked Language Modelling (MLM), focusing on learning specific language embeddings. These embeddings can then be employed as input for task adapters or combined with AdapterFusion for extracting latent knowledge for downstream tasks. 

\subsubsection{AdapterFusion}
Language adapters are introduced to extract language-specific embeddings from the internal structure of an LM based on an abstract objective function, such as MLM, to learn the general representations of a language. AdapterFusion aims to extract and compose the latent knowledge from multiple language adapters for a downstream task such as code summarization. For example, given a set of $N$ language adapters, the output of adapterFusion is a weighted sum of outputs from the language adapters, while the weights of the LMs ( $\Theta$) and the language adapters $(\theta_1,...,\theta_N)$ are fixed. This is shown in Equation \ref{eq:fusion-adapter}:
\begin{equation}
\Phi= \textrm{argmin } L(D;\Theta,\theta_1,...,\theta_N)
\label{eq:fusion-adapter}
\end{equation}

where $\Phi$ consists of the $Key_l$, $Value_l$ and $Query_l$ metrics at each layer $l$. At each Transformer block, the output of the feed-forward sub-layer is taken to be the $Query$, and the output of each language adapter is used for both $Key$ and $Value$ vectors.
Figure~\ref{fig:fusion-internal} shows the internal structure of AdapterFusion.

\section{Adversarial Fusion Adapter}
\label{section:approach}
In this section, we describe the architecture of our approach, AdvFusion, before proposing a learning algorithm for it. 

\subsection{Architecture}
\label{subsec:architecture}

\begin{figure}
\centering
    \scalebox{.5}{
    \includegraphics{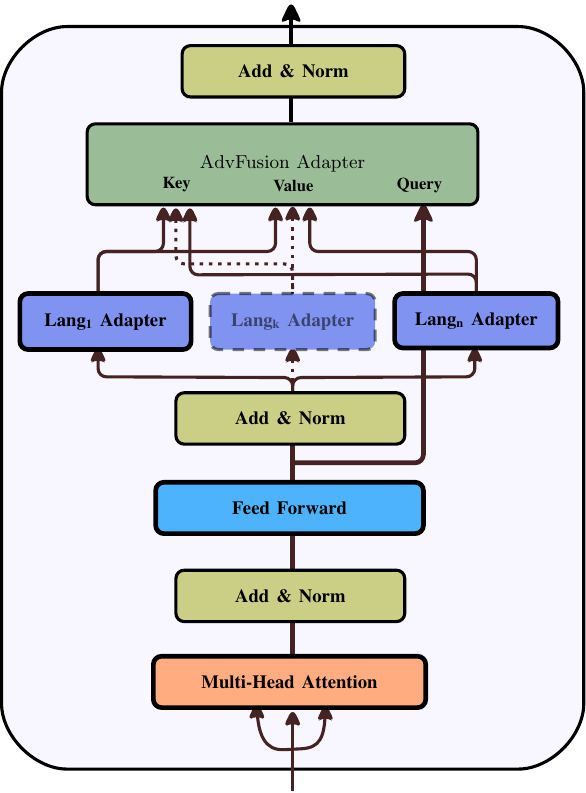}
    }
\caption{The adversarial training phase of AdvFusion. }
\label{fig:advfusionphase1}
\end{figure}

\begin{figure}
\centering
    \scalebox{.5}{
    \includegraphics{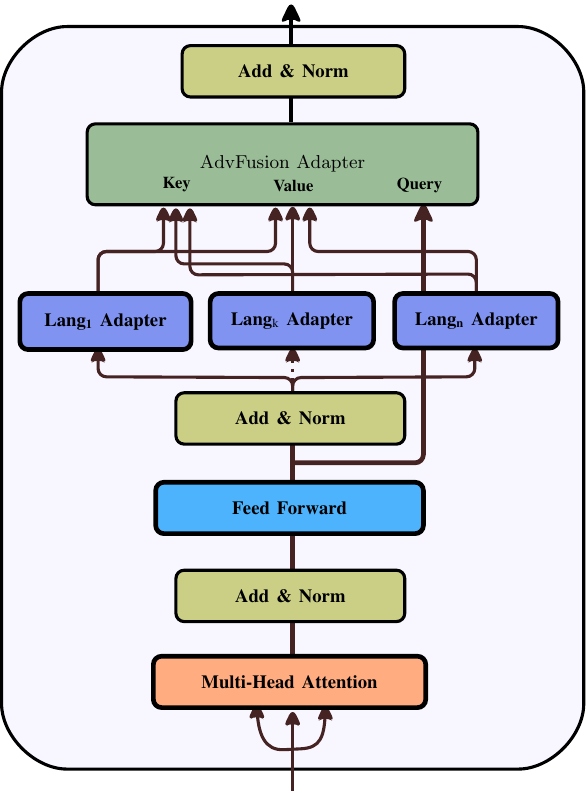}
    }
\caption{ The fine-tuning phase of AdvFusion. }
\label{fig:advfusionphase2}
\end{figure}

AdapterFusion can leverage the language adapter corresponding to the language of the current input better \cite{pfeiffer2020adapterfusion}, i.e., it pays more attention to the language adapter of the target task. This is mainly due to its internal attention mechanism.  
This mechanism prevents the effective utilization of the other language adapters, thus rendering them redundant. 
In light of this, we propose a new architecture, AdvFusion, that requires AdapterFusion to learn more from the other language adapters that are trained using a different language from the target task. 
Our approach consists of two training phases, the Adversarial training phase and the Fine-tuning phase: 
\begin{enumerate}

    \item 
    Adversarial training phase (see Fig.~\ref{fig:advfusionphase1}): In this phase, (i) the weights of the language adapter that corresponds to the language of the target task are set to zero, while (ii) the weights of the code-LM and the other language adapters are fixed. Then, (iii) AdvFusion is trained on the entire dataset. This phase allows AdvFusion to learn from the other language adapters.
    \label{phase1}
    \item
    Fine-tuning phase (see Fig.~\ref{fig:advfusionphase2}): In this phase, AdvFusion would have learnt from the other language adapters in the earlier phase. However, we still want AdvFusion to learn from the language adapter that corresponds to the language of the target task. Thus, (i) we restore the trained weights of the language adapter that corresponds to the language of the target task, while still (ii) fixing the weights of the code-LM and all language adapters. Then, (iii) the weights of AdvFusion are fine-tuned.
    \label{phase2}

\end{enumerate}

\subsection{Learning Algorithm}
\label{subsec:learning}
In this section, we formalize the learning procedure of AdvFusions. Let $\Theta$ and $\theta_i$ denote the parameters of the code-LM and each language adapter, $language_i$, respectively. We introduce the $ \Psi$ parameters to learn an embedding space from $N$ language adapters for a downstream task. For the adversarial training phase, we formalize it as follows:
\begin{equation}
\begin{aligned}
\Psi \leftarrow \mathop{\textrm{argmin }}_{\Psi}  
 \sum_{m=1}^{N} L(D_{m}; \Theta, \theta_{1},..,\theta_{m-1},\theta_{m+1},..,\theta_{N},\Psi)
\end{aligned}
\label{eq:adv-phase1}
\end{equation}

where L is the loss function of the downstream task, and $D_m$ denotes the $language_m$ dataset. In this step, AdvFusion learns to compose the embeddings of $N-1$ language adapters at each training step (recall that we are only interested in learning from the other language adapters that differ from the language of the target task in the adversarial training phase, thus we are only learning from $N-1$ language adapters). 

In the second phase, we employ all the language adapters to train the $\Psi$ parameters as follows:

\begin{equation}
\begin{aligned}
\Psi \leftarrow \mathop{\textrm{argmin }}_{\Psi}  
 \sum_{m=1}^{N} L(D_{m}; \Theta, \theta_{1},..,\theta_{N},\Psi)
\end{aligned}
\label{eq:adv-phase2}
\end{equation}

As illustrated in Fig. \ref{fig:advfusionphase1}, $ \Psi$ consists of the \textit{Key}, \textit{Value} and \textit{Query} parameters, denoted by $K_l$, $V_l$ and $Q_l$ at the Transformer layer $l$, respectively. 

Let $h_l$ denote the output of the feed-forward sub-component at the Transformer layer $l$. This is an input to AdvFusion. The output of the language adapter $i$ at the Transformer layer $l$, denoted as $z_{l,i}$, is the input for both the \textit{Key} and \textit{Value} transformations at the Transformer layer $l$. We compute the output of AdvFusion, denoted by $O_l$, as follows:

\begin{equation}
\begin{aligned}
& S_{l} = \textrm{softmax}(h^T_l Q_l \otimes z^T_{l,n} K_l) \\
& z'_{l,n} = z^T_{l,n} V_l \\
& Z'_l = [z'_{l,0},...,z'_{l,N}] \\
& O_l = S^T_l Z'_l
\label{eq:adv-phase-output}
\end{aligned}
\end{equation}

Given the embeddings of each language adapter ($z_n$), AdvFusion learns a weighted mixer of the available trained language adapters. In equation \ref{eq:adv-phase-output}, $\otimes$ represents the dot product and $n$ refers to two different things in each of the phases in AdvFusion. In the adversarial training phase (first phase),  $n \in \{1,...,m-1,m+1,...,N\}$ while in the fine-tuning phase (second phase), $n \in \{1,...,N\}$.

\section{AdvFusion on Code-Language Models}
\label{section:advfusion-setup-on-codelms}

In this section, we summarize our experiments and findings from our original paper on AdvFusion, where most writings are adopted from~\cite{saberi2025advfusion}. 

\subsection{Setup}

\paragraph{Backbone Models} 
Studies on how code language models and code large language models understand code reveal that fine-tuning smaller models on a target task could be more effective as compared to code-LLMs with billions of parameters~\cite{ma2024unveiling}. This finding is further supported by Dou et al.~\cite{dou2023towards} for another software engineering task. Given these findings, we have selected CodeT5+(220M)~\cite{wang2023codet5+}, CodeBERT~\cite{feng2020codebert} and GraphCodeBERT\cite{guo2020graphcodebert} as our baseline models. CodeT5+220M is considered an improved version of CodeT5~\cite{wang2021codet5}. The other models have been extensively researched in the field of software engineering~\cite{ahmed2021multilingual,saberi2023model,lu2021codexglue,wang2023oneAdapter,chung2014empirical}.
Additionally, these models are studied for multilingual fine-tuning for these two tasks and therefore serve as a basis in our comparisons~\cite{ahmed2021multilingual}.

\textit{CodeT5+} is an advanced code-LM designed by Wang et al. \cite{wang2023codet5+} to overcome limitations in existing code models, which often rely on rigid encoder-only or decoder-only architectures. It introduces a flexible, modular approach, allowing customization for various code-related tasks. CodeT5+ achieves superior performance compared to other code-LMs of similar size by incorporating a combination of pre-training techniques, including span denoising and contrastive learning.

\textit{CodeBERT}, as introduced by Feng et al. \cite{feng2020codebert}, is a bimodal pretrained model designed for both natural language and programming language understanding. Its architecture is based on Transformers. CodeBERT employs two pre-training objectives, namely Masked Language Modelling and Replaced Token Detection. These objectives are specifically chosen to enhance its capabilities in supporting tasks such as code search and code documentation generation.

\textit{GraphCodeBERT}, introduced by Guo et al. \cite{guo2020graphcodebert}, is a pioneering pre-trained model designed to enhance code comprehension tasks such as code summarization. GraphCodeBERT utilizes semantic-level information from code, specifically focusing on aspects like data flow. This pre-training approach employs a 12-layer transformer-based architecture. It is pre-trained on Masked Language Modelling, Edge Prediction and Node Alignment objective functions.

\paragraph{Tasks and Metrics}
We study the two tasks of code summarization and method name prediction. 

\textit{Code Summarization.} 
Given a code snippet, the task of code summarization is to describe its functionality. 
{It enhances code readability, aids in program comprehension, and facilitates easier maintenance and documentation. By providing summaries, developers can quickly understand the purpose and functionality of a piece of code without delving into its implementation details \cite{nie2022impact}.}
Code summarization is chosen as it is a widely studied task, and the effects of multilingual fine-tuning for this task are investigated in previous research \cite{ahmed2021multilingual,wang2023oneAdapter}. 

We evaluate the code summarization task using smooth-BLEU-4 \cite{papineni2002bleu}, which is a widely used metric in natural language generation tasks and many software engineering studies \cite{feng2020codebert,guo2020graphcodebert,wang2021codet5,wang2023oneAdapter,tang2020multilingual}.
BLEU is a precision-based metric that measures the n-gram geometric precision between the generated summary (i.e., n-gram hit) and the ground truth summary (i.e., total n-gram count) \cite{papineni2002bleu}.

\textit{Method Name Prediction.} 
The objective of the method name prediction task is to generate the most fitting method name that describes the purpose and functionality of the method's code. This task is chosen because naming methods accurately is crucial for code readability, maintainability, and understanding.

We report precision, recall and F1-score for the generated method names. F1 Score is the weighted average of Precision and Recall: $F1 = \frac{2 \cdot (P \cdot R)}{P + R}$. 
Where P and R stand for Precision and Recall, respectively.

Precision is computed as $P = \frac{TP}{TP+FP}$, whereas Recall is calculated as $R = \frac{TP}{TP+FN}$. P is calculated as the length of the intersection of ground truth tokens and generated output tokens (i.e., TP) divided by the length of output tokens (i.e., TP + FP). Similarly, R represents the recall, calculated as the length of the intersection of ground truth tokens and generated output tokens (i.e., TP) divided by the number of ground truth tokens (i.e., TP + FN).

\paragraph{Baselines}
AdvFusion in a model should be compared against the same model+AdapterFusion. For example, we should compare CodeBERT+AdvFusion with CodeBERT+AdapterFusion. 
To show the effectiveness of the base PEFT architecture we used, we also provide the results for mono-lingual fine-tuning, including model+TaskAdapters and model+LoRA~\cite{hu2021lora}. 
Note that we perform experiments on LoRA \cite{hu2021lora} as it is a widely used PEFT method. This enables us to compare its performance against other approaches and AdvFusion.

\paragraph{Training Details}
To train AdvFusion, in the first phase, we (1) fix the weights of the language adapter, (2) temporarily set the weights of the language adapter corresponding to the current input (i.e., the language of the target task) to zero, and (3) train the weights of AdvFusion on our target task. In the second phase, we restore the weights of the language adapter corresponding to the input and allow AdvFusion to learn from the language adapter that corresponds to the language of the current input.

Moreover, we evaluate the contribution of each programming language to a target programming language. Here, we choose Ruby, as it is named as a low-resource language in previous studies~\cite{chen2022transferability}, and it has been shown that it can benefit from other languages. 
For this purpose, we compute the contributions by feeding the Ruby test dataset into CodeBERT+AdapterFusion. Then, we aggregate the attention scores from each language adapter in each layer, normalize them (i.e., min-max normalization), and obtain the percentage of each language's contribution. We repeat these steps for CodeBERT+AdvFusion to compare its ability with AdapterFusion in extracting knowledge from other programming languages for Ruby. You can find the other language contributions on the repository page\footnote{https://github.com/ist1373/AdvFusion}. All experiments are conducted on an Nvidia Tesla V100 32GB GPU.

\paragraph{Datasets}
As Pfeiffer et al. have performed an extensive hyperparameter search over adapters, we use their reported optimal settings in our adapters' hyperparameters \cite{pfeiffer2020adapterfusion}.
We use the CodeSearchNet dataset \cite{husain2019codesearchnet} for training the language adapters. It consists of datasets from 6 programming languages, and the size of each language is shown in Table~\ref{table:plm}. We train language adapters using Mask Language Modelling.
We fine-tune AdapterFusion and AdvFusion adapters on the CodeSearchNet dataset using the next token prediction objective function for code summarization.
For the method name prediction task, we exclusively utilize the code portion of the CodeSearchNet dataset. We then mask the method names and let each approach suggest new method names using the next token generation objective function.

\begin{table}[h!]
\centering
   
    \begin{tabular}{|c | c | c| c|} 
        \hline
        Language & \# of Bimodal Data & Language & \# of Bimodal Data \\ [0.5ex] 
        \hline
         Ruby  & 24,927  & Python  & 251,820\\
         JavaScript  & 58,025  & Java  & 164,923\\
         Go  & 167,288  & PHP  & 241,241 \\ 
         \hline
    \end{tabular}
    \caption{Dataset statistics for Code Summarization and Method Name Prediction.  \cite{husain2019codesearchnet}}
    \label{table:plm}
  
\end{table}

\subsection{Results}

In this section, we present the results of our experiments to find i) whether using AdvFusion leads to a performance improvement in multilingual fine-tuning, and ii) quantify the attention that is placed on the target language from the other (non-target) languages in AdvFusion. 

\subsubsection{Performance of Multilingual PEFT with AdvFusion}

We evaluate how much improvement we could gain by using other programming languages; therefore, transferring knowledge in the multilingual parameter-efficient fine-tuning of Code-LMs. 
In Table \ref{table:code-summary}, we present the BLEU scores for both multilingual and monolingual PEFT approaches applied to Code-LMs. The multilingual approaches include Code-LM with AdvFusion and AdapterFusion, while the monolingual approaches involve Code-LM with TaskAdapter and LoRA. Although the base Code-LMs are the same, the key difference lies in the fine-tuning strategies used.

\begin{table*}[t!]
\centering

\begin{tabular}{|l | c | c| c| c| c| c|} 
 \hline
 \textbf{Models} & \textbf{Ruby} & \textbf{JavaScript} & \textbf{Go} & \textbf{Python} & \textbf{Java} & \textbf{PHP} \\ [0.5ex] 
  \hline

 CodeT5p+AdvFusion & 14.70 & \textbf{14.96}  & 18.25  & \textbf{18.98} & \textbf{18.78}  & \textbf{23.87}  \\ 
 CodeT5p+AdapterFusion & \textbf{14.79} & 14.82  & 18.30 & 18.94  & 18.71  & 23.80 \\ 

 CodeT5p+TaskAdapter & 13.99 & 14.31  & \textbf{18.34} & 18.91  & 18.68  & 23.71 \\ 

 CodeT5p+LoRA & 13.56 & 14.25  & 18.08 & 18.88 & 18.67  & 23.47 \\ 
\rowcolor{gray!30}
  CodeT5p (FFT)  & 14.55 & 15.16  & 19.00 & 19.77 & 19.60  & 25.13 \\ 
 \hline
 
 GraphCodeBERT+AdvFusion & \textbf{16.47} & \textbf{15.89}  & \textbf{19.96}  & 18.49 & \textbf{18.97}  & \textbf{24.83}  \\ 
 GraphCodeBERT+AdapterFusion & 15.57 & 14.49  & 18.21 & 17.86  & 18.21  & 23.54 \\ 

 GraphCodeBERT+TaskAdapter & 14.39 & 14.53  & 18.47 & 17.88  & 17.29  & 23.36 \\ 

 GraphCodeBERT+LoRA & 14.48 & 14.63  & 17.8 & \textbf{18.50}  & 17.16  & 24.13 \\ 

\rowcolor{gray!30}
GraphCodeBERT (FFT)  & 12.62 & 14.79 &18.40 & 18.02 & 19.22 & 25.45\\ 
 \hline
 CodeBERT+AdvFusion & \textbf{16.53} & \textbf{16.80} & \textbf{19.69} & 18.28 & \textbf{19.94} &25.20  \\ 
 CodeBERT+AdapterFusion & 15.38 & 15.88 & 18.31 & 18.40 & 19.04 & 25.17 \\ 

 CodeBERT+TaskAdapter & 14.12 & 15.67 &18.51 & \textbf{18.47} & 18.99 & \textbf{25.55} \\ 

 CodeBERT+LoRA & 12.27 & 13.67  &19.01 & 17.07 & 16.58 & 23.08 \\ 
 \rowcolor{gray!30}
 CodeBERT(FFT) & 12.16 & 14.90 &18.07 & 19.06 & 17.65 & 25.16 \\ 
\hline
 
 \hline
\end{tabular}

\caption{Smooth BLEU-4 scores on code summarization. When AdvFusion is combined with Code-LMs, we saw an improved performance in the majority of the datasets. FFT stands for Full Fine-Tuned. }
\label{table:code-summary}
\end{table*}

With CodeBERT+AdvFusion and GraphCodeBERT+AdvFusion, we observe improvements for Ruby, JavaScript, Go, and Java.
However, for Python and PHP, CodeBERT+TaskAdapter and GraphCodeBERT+LoRA show higher performance. We attribute this to the larger training data available for Python and PHP compared to Ruby and JavaScript, which have fewer samples. The smaller datasets for Ruby and JavaScript suggest that these languages still benefit from additional knowledge transfer.

We also compare the performance of AdvFusion with the state-of-the-art PEFT method, LoRA. In five of the programming languages evaluated (excluding Python), AdvFusion consistently outperforms LoRA. Performance gains are especially pronounced for CodeBERT and GraphCodeBERT, while the improvement for CodeT5p is less substantial. To better understand this discrepancy, we manually analyzed the outputs of CodeT5p+AdapterFusion and CodeT5p+AdvFusion against the ground truth targets, as shown in Table \ref{table:manual_comparison}. Although the overall improvement for CodeT5p is modest, our analysis reveals that AdvFusion tends to capture finer details more effectively.

In terms of parameter efficiency, both AdapterFusion and AdvFusion are more efficient than fully fine-tuning CodeBERT.
As shown in Table \ref{table:adv-time}, the average time to fine-tune all  CodeBERT weights was approximately 8 hours. In contrast, fine-tuning CodeBERT with AdvFusion took approximately 5.5 hours, representing a reduction of about 44\% in training time compared to the full fine-tuning of the entire model.

\begin{table}
    \centering
    \scalebox{.95}{
    \begin{tabular}{c|c|c|c}
    \hline
         Language & CodeBERT & CodeBERT+AdvFusion & Time reduction\\
         \hline
         Ruby &  492 & 328 & -33\% \textcolor{Gr}{$\downarrow$}\\ 
         JavaScript &  493 & 344 & -30\% \textcolor{Gr}{$\downarrow$}\\ 
         Go  &   511 & 336 & -34\% \textcolor{Gr}{$\downarrow$}\\ 
         Python  & 493 & 323 & -34\% \textcolor{Gr}{$\downarrow$}\\ 
         Java &  494 & 341 & -31\% \textcolor{Gr}{$\downarrow$}\\ 
         PHP &  506 & 338 & -33\% \textcolor{Gr}{$\downarrow$}\\
         \hline
    \end{tabular}
    }
    \caption{AdvFusion time efficiency for code summarization. Numbers represent training time in minutes, with the last column showing percentage improvement. Times reflect training for 20,000 training steps.}
    \label{table:adv-time}
\end{table}

We perform method name prediction on our baseline CodeLMs. The results are shown in Table~\ref{table:mnp}.
For this task, we observe that both AdapterFusion and AdvFusion have a larger impact on the results when they are added to GraphCodeBERT. This improvement is significant for all languages. 
For both models, AdvFusion slightly improves the results of AdapterFusion or achieves the same scores. We hypothesize that the variation could stem from the initial disparity in inputs and training methods between CodeBERT and GraphCodeBERT. GraphCodeBERT, utilizing dataflow graphs as input, gains a deeper understanding of the internal connections within code elements. This enhanced comprehension of the relationships among the programming languages enables GraphCodeBERT to suggest more effective method names by leveraging the knowledge from other programming languages for the language of the target task when AdvFusion is used.

\begin{table*}[!ht]
\centering
\begin{adjustbox}{width=1\textwidth,center}
\begin{tabular}{@{}cccccccccccccccccccccccc@{}}
 \hline
\textbf{Model} &\multicolumn{3}{c}{\textbf{Ruby}}  &  & \multicolumn{3}{c}{\textbf{Javascript}} &  & \multicolumn{3}{c}{\textbf{Go}} &  & \multicolumn{3}{c}{\textbf{Python}} &  & \multicolumn{3}{c}{\textbf{Java}} &  & \multicolumn{3}{c}{\textbf{PHP}}\\ \cline{2-4}\cline{6-8} \cline{10-12}\cline{14-16} \cline{18-20} \cline{22-24}
&  $P$  & $R$ & $F1$ &  & $P$ & $R$ & $F1$ &  & $P$ & $R$ & $F1$ &  & $P$ & $R$ & $F1$ &  & $P$ & $R$ & $F1$ &  & $P$ & $R$ & $F1$  \\ [1em] 

\specialcell{CodeT5p + \\AdvFusion} &  \textbf{0.55}   &   \textbf{0.55}    &   \textbf{0.55}    &  &   \textbf{0.59}    &   \textbf{0.56}  &   \textbf{0.58}    
& &  \textbf{0.58}   &   \textbf{0.56}  &  \textbf{0.57}   &  &   0.61  &  0.61  &  0.61  & &  \textbf{0.61}   & 0.57   &  \textbf{0.59}   & &  \textbf{0.49}  &  0.46  &  \textbf{0.48} \\ [1em]

\specialcell{CodeT5p + \\AdapterFusion} &  0.54   &   0.54    &   0.54    &  &   0.57    &   0.55  &   0.56    
& &  0.55   &   0.53  &  0.54   &  &   0.60  &  0.59  &  0.60  & &  0.59   & 0.56   &  0.57   & &  0.47  &  0.44  &  0.46 \\ [1em]

\specialcell{CodeT5p + \\TaskAdapter} &  0.53   &   0.54    &   0.54    &  &   0.54    &   0.57  &   0.55    
& &  0.55   &   0.57  &  0.56   &  &   0.61  &  0.61  &  0.61  & &  0.60   & 0.57   &  0.59   & &  0.48  &  0.46  &  0.47 \\ [1em]

CodeT5p+LoRA &  0.53   &   0.52    &   0.53    &  &   0.53    &   0.56  &   0.55    
& &  0.54   &   0.56  &  0.55   &  &   \textbf{0.61}  &  \textbf{0.61}  &  \textbf{0.61}  & &  0.57   & \textbf{0.59}   &  0.58   & &  0.48  &  0.45  &  0.46 \\ [1em]

\hline

\specialcell{CodeBERT + \\AdvFusion} &  \textbf{0.39}   &   0.32    &   \textbf{0.35}    &  &   0.19    &   0.14  &   0.16    
& &  \textbf{0.46}   &   \textbf{0.46}  &  \textbf{0.45}   &  &   \textbf{0.47}  &  \textbf{0.45}  &  \textbf{0.46}  & &  0.43   & 0.34   &  0.37   & &  \textbf{0.45}  &  \textbf{0.43}  &  \textbf{0.44} \\ [1em]

\specialcell{CodeBERT + \\AdapterFusion} &  0.38   &   0.30    &   0.32    &  &   0.19    &   0.14  &   0.16    
& &  0.45   &   0.40  &  0.41   &  &   0.44  &  0.34  &  0.37  & &  0.43   & 0.34   &  0.37   & &  0.45  &  0.38  &  0.40 \\ [1em]

\specialcell{CodeBERT + \\TaskAdapter} &  0.35   &   0.30    &   0.30    &  &   0.19    &   0.14  &   0.16    
& &  0.45   &   0.40  &  0.41   &  &   0.44  &  0.34  &  0.37  & &  0.43   & 0.34   &  0.37   & &  0.45  &  0.38  &  0.40 \\ [1em]

CodeBERT+LoRA &  0.36   &   \textbf{0.33}    &   0.34    &  &   \textbf{0.21}    &   \textbf{0.16}  &   \textbf{0.18}    
& &  0.45   &   0.42  &  0.43   &  &   0.43  &  0.45  &  0.44  & &  0.42   & \textbf{0.40}   &  \textbf{0.41}   & &  0.41  &  0.44  &  0.43 \\ [1em]

\hline

 \textbf{\specialcell{ Graph\\CodeBERT + \\AdvFusion}} &  \textbf{0.42}   &   \textbf{0.32}    &   \textbf{0.36}    &  &   \textbf{0.58}    &   \textbf{0.58}  &   \textbf{0.58}   
& &  \textbf{0.51}   &   \textbf{0.51}  &  \textbf{0.51}   &  &   0.49  &  0.40  & 0.42  & &  \textbf{0.52}   &  \textbf{0.50}   &  \textbf{0.51}   & &  \textbf{0.54}  &  \textbf{0.53}  &  \textbf{0.54} \\[1em]

 \textbf{\specialcell{ Graph\\CodeBERT + \\AdapterFusion}} &  0.40   &   0.30    &   0.35    &  &   0.57    &   0.57  &   0.57   
& &  0.48   &   0.49  &  0.47   &  &   0.48  &  0.38  & 0.41  & &  0.48   &  0.49   &  0.48   & &  0.52  &  0.50  &  0.51 \\[1em]

 \specialcell{Graph\\CodeBERT + \\TaskAdapter} &  0.40   &   0.33    &   0.35    &  &   0.24    &   0.22  &   0.23    
& &  0.47   &   0.42  &  0.43   &  &   0.47  &  0.38  & 0.40  & &  0.45   &  0.37   &  0.40   & &  0.48  &  0.41  &  0.43 \\[1em]

 \specialcell{Graph\\CodeBERT+LoRA} &  0.39   &   0.32    &   0.35    &  &   0.28    &   0.24  &   0.26    
& &  0.51   &   0.45  &  0.47   &  &   \textbf{0.50}  &  \textbf{0.44}  & \textbf{0.45}  & &  0.48   & 0.43   &  0.44  & &  0.49  &  0.45  &  0.46 \\[1em]

 \bottomrule
\end{tabular}
\end{adjustbox}
\caption{
The Precision (P), Recall (R), and F1-Score (F1) metrics were assessed on each programming language across various settings. When AdvFusion is combined with Code-LMs, we saw an improved performance in the majority of the datasets.}
\label{table:mnp}
\end{table*}

\vspace{10pt}
\begin{mdframed} [backgroundcolor=gray!20, linewidth=1pt]
\textbf{When \textit{AdvFusion} is used for fine-tuning Code-LMs, we can achieve better or on par results compared to other PEFT methods. The improvement is observed more for the three programming languages Ruby, JavaScript, and Go for code summarization. As AdvFusion is a PEFT method, the training time is reduced, and approximately 80\% fewer parameters are trained compared to full fine-tuning Code-LMs.}
\end{mdframed}

\begin{table*}[h!]
\centering
\begin{adjustbox}{width=1\textwidth,center}
\begin{tabular}{|c|c|c|c|}
\hline
\textbf{Samples} & \textbf{CodeT5p+Fusion}  & \textbf{CodeT5p+AdvFusion} & \textbf{Target} \\ \hline
sample 1(Javascript) & Parse a segment \colorbox{red!30}{of a string}. & Parse a segment into \colorbox{green!30}{a single object}. & Parse a segment and convert it into json. \\ \hline

sample 2(Javascript) & Transform a metadata object into a \colorbox{red!30}{ string}. & \begin{tabular}{@{}c@{}} Transform a metadata object \\  into a \colorbox{green!30}{list of tokens}. \end{tabular}  & \begin{tabular}{@{}c@{}} Transform token names to formats expected \\ by Sassdoc for descriptions and aliases \end{tabular} \\ \hline

sample 3(PHP) & Create a new \colorbox{red!30}{model}. & Create a new \colorbox{green!30}{database table}. & Store new database table. \\ \hline

sample 4(PHP) & \colorbox{red!30}{Deletes all files} in the media picker table. & \colorbox{green!30}{Cleanup}  the media picker data & \begin{tabular}{@{}c@{}} Remove translations \\ images and files related to a BREAD item.\end{tabular} \\ \hline

sample 5(Go) & \begin{tabular}{@{}c@{}} Percentiles returns the percentage of \\ the given \colorbox{red!30}{number of elements}. \end{tabular}&  \begin{tabular}{@{}c@{}} Percentiles returns the percentage \\ of the given \colorbox{green!30}{array of floats}. \end{tabular}& \begin{tabular}{@{}c@{}}Percentiles returns percentile \\ distribution of float64 slice.\end{tabular} \\ \hline

sample 6(Go) & newPipelineHandler creates a new \colorbox{red!30}{pipeline handler}. & \begin{tabular}{@{}c@{}} newPipelineHandler returns a new \colorbox{green!30}{http}\\ \colorbox{green!30}{Handler that will handle the pipeline request}. \end{tabular} & \begin{tabular}{@{}c@{}c@{}c@{}} newPipelineHandler returns a handler for handling \\ raft messages from pipeline for RaftPrefix.\\ The handler reads out the raft message from request body \\and forwards it to the given raft state machine for processing.\end{tabular} \\ \hline
\end{tabular}
\end{adjustbox}

\caption{Comparison between CodeT5p+Fusion and CodeT5p+AdvFusion outputs with their ground truth. Samples are selected from the test set results of the CodeSearchNet dataset.}
\label{table:manual_comparison}
\end{table*}

\subsubsection{Languages' Contribution for a target Programming Language}

We assess the contribution of each language adapter across all programming languages for code summarization, comparing AdvFusion with AdapterFusion. Due to space constraints, we present only the results for Ruby, as the behaviour of other languages follows a similar trend. Figures illustrating the contributions of the other languages are available in the supplementary materials.

\begin{figure}

\includegraphics[width=\columnwidth]{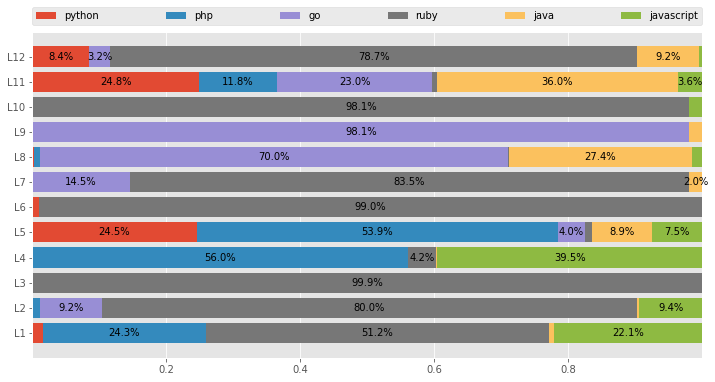}
\centering
\caption{The attention contribution from each programming language at each layer when we feed the Ruby test dataset to the fine-tuned AdapterFusion model.}
\label{fig:fusion-ruby}
\end{figure}

We extract the attention at AdvFusion and AdapterFusion when we fine-tune  CodeLMs+AdvFusion and CodeLMs+AdapterFusion, respectively (separate experiments). 
Figure~\ref{fig:fusion-ruby} demonstrates the contribution of each language at each layer in CodeBERT+AdapterFusion when the Ruby test dataset is fed to the fine-tuned model. 
It is noted that in most layers, a high percentage of attention (more than 80\%) is towards Ruby (the gray bar), rather than attending to other languages. In other words, not much is learned from other programming languages.

\begin{figure}
\includegraphics[width=\columnwidth]{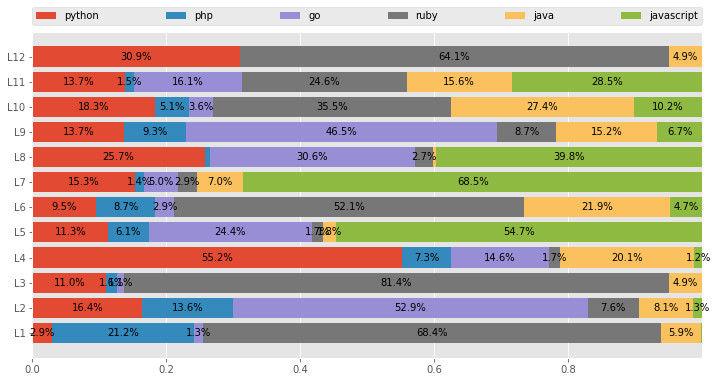}
\centering
\caption{The attention contribution from each programming language at each layer when we feed the Ruby test dataset to the fine-tuned AdvFusion model.}
\label{fig:adv-fusion-ruby}
\end{figure}

Figure~\ref{fig:adv-fusion-ruby} shows the contribution of each language in CodeBERT + AdvFusion when the Ruby test dataset is fed to the fine-tuned model. The y-axis is the layer number in CodeBERT, and the x-axis shows the percentage of contribution of each language. Here, AdvFusion pays more attention to other programming languages. 
For instance, Ruby has the following learning: it learns more from Go in the second layer (i.e., $52.9\%$ of attention is grabbed from the Go adapter), it learns more from Python than Ruby in the fourth layer (i.e., $56.2\%$), and it learns more from JavaScript in layer seven. Even in the higher layers, learning from other languages is continued and the attention is distributed to other languages, and not only focused on Ruby. 
More interestingly, PHP is the most resourceful language in the dataset, but its contribution to Ruby is less than other languages. This suggests that there is no relationship between the size of the language dataset and its contribution to Ruby.

\vspace{3mm}

\begin{mdframed} [backgroundcolor=gray!20, linewidth=1pt]
\textbf{Programming languages could benefit from the other resourceful languages differently in different layers. Higher-resource languages do not necessarily contribute more to the low-resource language, such as Ruby.}
\end{mdframed}

\subsection{When to Use AdvFusion on Pre-Trained Code Language Models?}

\textbf{\textit{When should we use adapters for monolingual fine-tuning? } }

In our experiments, we found that adapter-based fine-tuning is as effective as standard fine-tuning for high-resource languages in code summarization, while being more computationally efficient. It also enhances results for low-resource languages. Low-resource languages are those that have less training data available. Hence, we recommend adapter-based fine-tuning for monolingual fine-tuning in code summarization.
This result is similar to the findings in the literature \cite{liu2023empirical,weyssow2023exploring}.
We also observe that for Code-LMs in our study, adapters perform better than LORA and are a better choice among these two PEFT approaches.

However, for method name prediction on languages with limited resources, employing task adapters can still yield benefits without significant performance decline, while also reducing memory and time in fine-tuning.

\textbf{\textit{When should we consider knowledge transfer in multilingual fine-tuning?}}

Multilingual fine-tuning, as shown by Ahmed et al. \cite{ahmed2021multilingual}, often outperforms monolingual fine-tuning across resource levels. Table \ref{table:code-summary} highlights that some languages, like PHP, benefit less from multilingual adapters (e.g., AdapterFusion, AdvFusion) compared to full fine-tuning, possibly due to limited cross-language utility or insufficient PEFT parameter capacity. Python and Java show mixed results with PEFT, while AdvFusion effectively improves performance for low-resource languages by leveraging insights from Ruby and others.

\textbf{\textit{Which languages could a low-resource language take advantage of in a multilingual setting?}}

We observed that when using AdvFusion, Ruby has benefited from Go, Python and JavaScript, as depicted in Figure \ref{fig:adv-fusion-ruby}. This study does not focus on the syntactic or semantic similarities between the source and target programming languages but rather on which languages are most useful for Ruby from the perspective of a model in practice. Continuation of other programming languages is provided in supplementary materials.

Figure~\ref{fig:adv-sample} represents a heatmap generated from a Ruby sample fed into CodeBERT + AdvFusion. 
The x-axis displays Ruby tokens, while the y-axis shows the six programming languages of the CodeSearchNet dataset. Lighter colours indicate higher attention. This heatmap illustrates the attention each token receives from each programming language in the dataset.

The highest attention on the tokens is from other language adapters than the Ruby adapter; as observed, the attention from the Ruby adapter is very low (note the Ruby adapter row, which is dark everywhere).
However, for instance, the function signature of the sample, \code{sum},\code{(},\code{a},\code{b},\code{)} received more attention from Go rather than Ruby, and also the document tokens corresponded to the function signature, \code{the}, \code{sum}, and \code{of}, are paid more attention by Go. 
This is aligned with our observations in Figure \ref{fig:adv-fusion-ruby}, as discussed in RQ2.

\textbf{\textit{When can adapters be helpful, in terms of architectures and tasks?}}

In our study, incorporating adapters into the CodeT5 baseline for code summarization led to a performance decline. We attribute this to the pre-existing decoder stack in CodeT5, which may limit adaptability compared to models like CodeBERT or GraphCodeBERT that train the decoder from scratch. A similar issue was reported in \cite{wang2023oneAdapter}, where CodeT5 fine-tuning underperformed relative to CodeBERT and GraphCodeBERT.

\textbf{\textit{Which target tasks could benefit from multilingual fine-tuning using AdvFusion? }}\\
We have conducted experiments on code summarization and method name prediction, demonstrating the effectiveness of AdvFusion. We hypothesize that other tasks with consistent output modalities across datasets could similarly benefit from AdvFusion and AdapterFusion architectures. For instance, tasks like code review and commit message generation—where the output is natural language—could leverage multilingual fine-tuning, provided there are datasets available in multiple programming languages.

\begin{figure}
\includegraphics[width=\columnwidth]{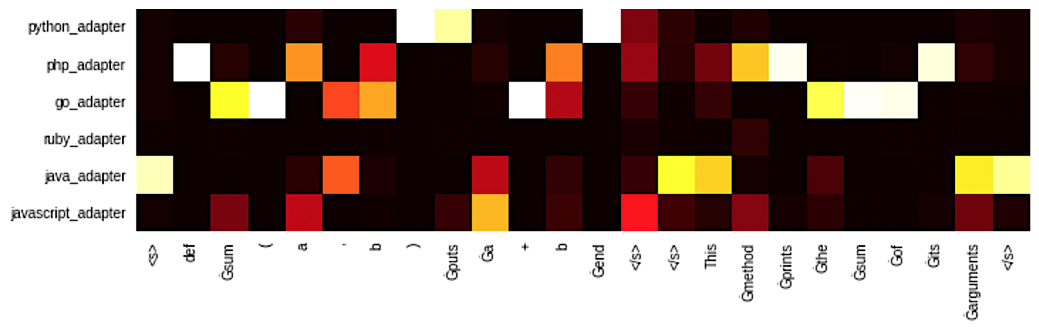}
\centering
\caption{AdvFusion's attention heatmap across six language adapters for a Ruby sample in the fine-tuned model. The X-axis displays code tokens, while the Y-axis shows attention from each adapter.}
\label{fig:adv-sample}
\end{figure}

\section{AdvFusion on Code-LLMs: Experimental Setup}
\label{section:advfusion-setup-on-codellms}
In this section, we explain the experiment setup and training details of applying AdvFusion on Code-LLMs. 

\subsection{Code-LLMs}

We conducted experiments on four popular and widely used open-source code-LLMs, including CodeLlama 7B, DeepSeek-Coder 1.3B, and Qwen2.5-Coder 1.5B and 3B.
Our goal in selecting these models was to cover a range of recent Code-LLM architectures and sizes while keeping computational requirements manageable, given the large number of experiments in this study. To this end, we focused on smaller models that are specialized for coding tasks and have demonstrated strong performance in software engineering benchmarks. We included DeepSeek-Coder 1.3B and Qwen2.5-Coder 1.5B as compact variants of state-of-the-art architectures, along with CodeLlama 7B (i.e., the latest code-specialized variant of the Llama family). To investigate the effect of model size within the same architecture, we also included Qwen2.5-Coder 3B. Overall, our selection aims to explore diverse architectures and include strong-performing code-specialized models, while maintaining feasible resource requirements.

\textit{CodeLlama} \cite{roziere2023code} is a model pre-trained on both general-purpose text and code data, achieving state-of-the-art performance among open-source models for code-related tasks. In our experiments, we used CodeLlama with 7B parameters.

\textit{DeepSeek-Coder} \cite{guo2024deepseek} is trained on 2 trillion tokens covering 87 programming languages, enabling it to achieve a broad and comprehensive understanding of code. In our study, we selected DeepSeek-Coder with 1.3B parameters for evaluation.

\textit{Qwen2.5-Coder} \cite{hui2024qwen2} is trained on extensive datasets and further fine-tuned on datasets specifically designed for coding tasks, demonstrating strong code generation capabilities while retaining general language and mathematical reasoning skills. For our experiments, we employed the 1.5B and 3B parameter versions of Qwen2.5-Coder.

\subsection{Tasks}

In this extension, we focus on three new target tasks: commit message generation, code generation, and code translation, which we believe more accurately reflect the challenges encountered in real-world software engineering scenarios. Commit message generation serves as a more demanding alternative to code summarization, requiring the model to analyze and understand long code diffs and express their intent through concise, meaningful messages. Code generation has become a central topic in software engineering due to its wide range of applications, from intelligent code assistants to automated development tools, and represents a natural fit for generative architectures. Code translation extends this challenge further, as the model must infer the intended functionality of a source program and accurately reproduce it in a target language, a capability essential for cross-language migration, code refactoring, and interoperability. 
Overall, these tasks span three modalities: natural language to code, code to natural language, and code to code, allowing a comprehensive evaluation of generative capabilities. 
Note that we did not apply these new tasks to the Code-LM models presented in Section~\ref{section:advfusion-setup-on-codelms} (i.e., CodeBERT, GraphCodeBERT, and CodeT5), given that they are significantly smaller and not sufficiently capable for complex generative tasks. We note that evaluating the performance on Code-LMs is not the focus of this paper -- our focus is on newer tasks and Code-LLMs.

\textit{\textbf{Commit Message Generation (CMG)}}. This task converts a diff (or change in code) into a concise natural language description that documents the intent of the change. CMG has been widely studied in software engineering because high-quality commit messages materially improve code comprehension and long-term project maintenance \cite{commitbert2021, comsum2021}. In this work, we evaluate multilingual adapter training and subsequent fusion for CMG on Code-LLMs.

\textit{\textbf{Evaluation metrics for CMG.}}
We report common lexical metrics used in the commit message generation task: BLEU (BLEU-4) and ROUGE-L \cite{commitbert2021, comsum2021, kadel2024}. 
BLEU (Bilingual Evaluation Understudy) \cite{papineni2002bleu} measures the precision of n-grams between the generated and reference texts, indicating how much of the model’s output overlaps with the ground truth. BLEU-4, in particular, evaluates up to 4-gram matches.
ROUGE-L (Recall-Oriented Understudy for Gisting Evaluation)~\cite{lin2004rouge} emphasizes recall by computing the longest common subsequence between the generated and reference texts.

\textit{\textbf{Code Generation.}} \label{sec:task4}
Given a natural language description of a method/function, the code generation task is to generate its corresponding code. We chose this task because code generation has been extensively studied and plays an important role in facilitating software development \cite{jiang2024survey}.

\textit{\textbf{Evaluation metrics for Code Generation.}}
Given the small size of the selected models, they are not effective in producing code, and the Pass@K results were almost zero. Therefore, we used BLEU (BLEU-4) and ROUGE-L as the evaluation metrics for the code generation task \cite{lu2021codexglue, wang2021codet5}.

\textit{\textbf{Code Translation.}}
This task involves translating a program of a source programming language to a program of a target programming language. Code translation has many practical use cases, such as migrating existing code bases to newer programming languages and reusing obsolete modules in projects~\cite{jain2015modernization, khadka2014professionals}. Code translation is more challenging given that, under the hood, to translate a program of a programming language to another programming language, the language model must capture the objective of the program and reimplement it in the target program.

\textit{\textbf{Evaluation metrics for Code Translation.}} 
Our main evaluation metric for code translation is Pass@k, which measures the functional correctness of generated programs by assessing the proportion of correct generations among the model’s $k$ attempts. This metric captures the model’s ability to produce correct and executable code rather than merely textually similar output. We report results for both Pass@1 and Pass@10 to reflect single-try accuracy and performance under multiple generation attempts. To measure Pass@k, we use the PolyHumanEval benchmark \cite{polyhe}, which contains multilingual coding problems based on the HumanEval benchmark \cite{humaneval-codex-Chen2021EvaluatingLL} along with their respective test suites for evaluation. Additionally, we report BLEU-4 and ROUGE-L scores as textual similarity metrics on the test split of the code translation dataset.

\subsection{Datasets}

\textit{\textbf{Commit Message Generation.}}
All CMG experiments used CommitPackFT as the main data source because it contains high-quality commit messages \cite{muennighoff2023octopack} and is used in several previous studies \cite{cassano2024can, aggarwal2025robust}. CommitPack is the raw large collection of Git commits ($\approx$4 TB) scraped from permissively-licensed GitHub repositories across $\approx$350 programming languages; CommitPackFT is a heavily filtered $\approx$2 GB subset (277 languages) containing commit messages that more closely resemble natural-language instructions (filters include multi-word messages, an imperative/uppercased verb at the start, removal of external references, and other quality checks) \cite{muennighoff2023octopack}. For our experiments, we focused on the following five programming languages as language adapters: {Swift, Scala, Rust, C, and Java}, and defined the three low-resource target languages as {Swift, Scala, Rust} for AdapterFusion/AdvFusion experiments \cite{muennighoff2023octopack}. 
Statistics for the languages used in CommitPackFT are stated in Table \ref{tab:commitpackft-stats}. As the original dataset does not include predefined train, development, or test splits, we split the dataset accordingly.

\begin{table}[htbp]
\centering
\caption{Dataset statistics of CommitPackFT \cite{muennighoff2023octopack} used for high-quality commit-messages.}
\label{tab:commitpackft-stats}
\begin{tabular}{lrrrr}
\toprule
\textbf{Language} &\textbf{Train} &\textbf{Validation} &\textbf{Test} &\textbf{Total} \\\midrule
Java &16,508 &2,063 &2,064 &20,635 \\
C &6,804 &850 &852 &8,506 \\
Scala &4,032 &504 &504 &5,040 \\
Swift &3,879 &484 &486 &4,849 \\
Rust &2,396 &299 &301 &2,996 \\
\bottomrule
\end{tabular}
\vspace{1ex}
\end{table}

\textit{\textbf{Code Generation.}}
Models are trained using the xCodeEval dataset \cite{khan-etal-2024-xcodeeval} for program synthesis, where the goal is to generate code that solves a given problem. The xCodeEval dataset covers a diverse set of programming languages, such as C, C\#, Kotlin, Rust, Go, JavaScript, Ruby, and PHP. Since PHP has the lowest number, we selected it as the target (low-resource) language for our code generation experiments. Table \ref{tab:code-generation-dataset} presents the statistics for the programming languages used in the xCodeEval dataset for code generation.

\begin{table}[]
\centering
\caption{Dataset statistics of xCodeEval \cite{khan-etal-2024-xcodeeval} across programming languages chosen for code generation training.}
\label{tab:code-generation-dataset}
\begin{tabular}{lcccc}
\toprule
\textbf{Language} & \textbf{Train} & \textbf{Validation} & \textbf{Test} & \textbf{Total} \\
\midrule
C      & 143,443 & 18,016 & 18,049 & 179,508 \\
C\#   & 63,678 & 7,933 & 8,070 & 79,681 \\
Kotlin   & 41,535 & 5,143 & 5,153 & 51,831 \\
Rust  & 24,640 & 3,046 & 3,046 & 30,732 \\
Go  & 20,637  & 2,566 & 2,550 & 25,753 \\
Javascript & 12,741  & 1,589 & 1,586 & 15,916 \\
Ruby &  12,277 & 1,525 & 1,534 & 15,336 \\
PHP & 5,106  & 615 & 613 & 6,334 \\
\bottomrule
\end{tabular}
\vspace{1ex}
\end{table}

\textit{\textbf{Code Translation.}}
The training dataset for code translation used in this study is derived from the {NicheTrans} split of CodeTransOcean \cite{yan2023codetransocean}, which contains translation pairs from eight popular to $37$ low-resource programming languages. For this task, samples with multiple source languages and a singular target language are combined together, forming a many-to-one relation, and finally deduped. We limit the source languages of the derived dataset to C++, C\#, Go, Java, PHP, Python and VB, and the target languages to Julia, Ruby, Scala and Swift. The statistic of the resulting dataset is presented in Table \ref{tab:code-translation-dataset}.

\begin{table}
\centering
\caption{Dataset statistics of the code translation dataset, derived from the \emph{NicheTrans} split of CodeTransOcean \cite{yan2023codetransocean}. Each target language split contains six source languages of C++, C\#, Go, Java, PHP, Python and VB.}
\begin{tabular}{lrrrr}
\toprule
\textbf{Language} &\textbf{Train} &\textbf{Validation} &\textbf{Test} &\textbf{Total} \\\midrule
Julia &4,502 &1,131 &2,288 &7,921 \\
Ruby &4,463 &967 &1,904 &7,334 \\
Scala &4,517 &812 &1,605 &6,934 \\
Swift &2,844 &292 &653 &3,789 \\
\bottomrule
\end{tabular}

\label{tab:code-translation-dataset}
\end{table}

\subsection{PEFT Methods}

\textit{\textbf{Bottleneck adapter}}
Bottleneck adapters introduce bottleneck feed-forward layers in each layer of a Transformer model \cite{houlsby2019parameter}. In the experiment, we trained a Bottleneck Adapter for each programming language, and then trained AdapterFusion and AdvFusion on top of them.

\textit{\textbf{LoRA}} 
LoRA is a popular and lightweight training technique which freezes the pretrained model weights and injects trainable rank decomposition matrices into layers \cite{hu2021lora}. We compare LoRA with Code-LLMs trained using Bottleneck adapter combined with AdapterFusion and AdvFusion, as well as Code-LLMs trained using Compacter combined with AdapterFusion and AdvFusion.

\textit{\textbf{Compacter}}
Compacter combines low-rank decomposition with parameterized hypercomplex multiplication layers to create compact adapters with minimal trainable parameters \cite{karimi2021compacter}. We also trained a Compacter for each programming language, and then trained AdapterFusion and AdvFusion on top of them.

\subsection{Experimental Design}

As the first stage, all non-fusion PEFT models (i.e., Bottleneck Adapters, Compacter, and LoRA) undergo standard fine-tuning for each downstream task and programming language, using the Causal Language Modelling (CLM) objective. The resulting models are then evaluated independently, and their performance is reported separately.

In the second stage, we reuse the pretrained Bottleneck Adapters and Compacter PEFT modules from stage one to train fusion-based PEFT models, AdapterFusion and AdvFusion. For AdapterFusion, the pretrained PEFT modules are inserted into the base model in a frozen state (i.e., their parameters remain fixed during training), and a fusion module is added to each layer. The model is then fine-tuned using the CLM objective following the same training procedure as in the first step.

For AdvFusion, training proceeds in two steps. In the first step, we again insert the pretrained PEFT modules in a frozen state along with fusion modules, but mask out the PEFT module corresponding to the target language. For instance, if the composing PEFT modules are trained on Julia, Ruby, Scala, and Swift, and the target language is Ruby, the Ruby module is masked. This encourages the fusion layers to learn to integrate and attend to features from other languages. Once the masked fine-tuning step is complete, we unmask the target PEFT module and continue training for an equal number of epochs as in the first step.

\textbf{TaskAdapters.} Unlike our initial study~\cite{saberi2025advfusion}, we use TaskAdapters instead of LanguageAdapters to train AdapterFusion and AdvFusion. This is a necessary change that reflects the architecture of the decoder-only Code-LLMs used in this study. In contrast to our initial study, where all models were encoder-only or encoder-decoder-based and pretrained with the Masked Language Modelling (MLM) objective, all Code-LLMs in this work are decoder-only models pretrained on the Causal Language Modelling (CLM) objective for next-token prediction. Because LanguageAdapters were originally designed to align with the MLM objective, their adaptation layers expect bidirectional contextual representations. Applying them to decoder-only models, which rely on strictly autoregressive token dependencies, leads to a mismatch in both the learning signal and representation flow. This mismatch disrupted the pretrained models’ generation behaviour and degraded their performance. TaskAdapters, in contrast, align naturally with the CLM objective and integrate seamlessly into the causal decoding architecture, making them more suitable for our setup.

\textbf{Hyperparameters.} For Commit Message Generation and Code Generation, we used a batch size of 16 for smaller models and 2 for larger models. For Code Translation, the batch size is 4 across all model sizes. LoRA is configured with rank $r=16$ and scaling factor $\alpha=16$, while Compacter used a PHM dimension of 4 for all downstream tasks. Bottleneck Adapters and Fusion layers adopt their default hyperparameters as specified in their original works. All base models were quantized to 4-bit precision to enable efficient training.

All experiments were conducted on NVIDIA A100 (40 GB) and H100 (80 GB) GPUs. We employed AdapterHub \cite{pfeiffer2020adapterhub} implementations for bottleneck adapters, Compacter, and Fusion, and Hugging Face PEFT \cite{peft-library} for LoRA.

\section{AdvFusion on Code-LLMs: Results}
\label{section:advfusion-results-on-codellms}

In this section, we present the results of our experiments and focus on answering these research questions: 

\textbf{RQ1. How well does AdvFusion perform on Code-LLMs?}\\
In this RQ, we compare the performance of AdvFusion with its base PEFT, AdapterFusion, to investigate whether Code-LLMs and Code-LMs have similar performance.  

\textbf{RQ2. Does replacing Bottleneck adapters with Compacter in the AdvFusion architecture impact the performance?}\\
Compacter has shown comparable results for low-resource languages in previous studies \cite{esmaeili2024empirical}. 
Therefore, in this RQ, we conduct experiments replacing the base adapter in Fusion PEFT architectures (i.e., AdapterFusion and AdvFusion) with Compacter and investigate whether this change would improve the performance of the models.

In the following, we present the results. First, we discuss the general trend observed for each task and then explicitly for each task, we answer the RQs.

\subsection{Commit Message Generation}
\label{sec:results-cmg}

Table~\ref{tab:cmg_bleu_rouge_scala} shows the BLEU-4 and ROUGE-L scores for the CMG task using Scala as the target language across four models. Among the target languages evaluated in CMG (Rust, Scala, Swift), Scala exhibits a representative trend that closely mirrors the aggregated cross-language behaviour. For this reason, we will focus primarily on Scala to illustrate and generalize the overall findings of CMG. In the table, \emph{AdvFusion} refers to adversarial fusion using Bottleneck adapters, while \emph{AdvFusion+Compacter} refers to adversarial fusion using Compacter adapters. The same distinction applies to \emph{AdapterFusion} and \emph{AdapterFusion+Compacter}. AdapterFusion achieves the highest overall scores with an average BLEU-4 of $22.63$ and ROUGE-L of $42.60$. LoRA and TaskAdapter follow closely (TaskAdapter: BLEU-4 $22.01$, ROUGE-L $40.05$; LoRA: BLEU-4 $21.95$, ROUGE-L $40.31$). 
AdvFusion follows these scores closely, with BLEU-4 of $21.75$ and ROUGE-L $39.76$. 
In contrast, Compacter and AdvFusion variant with Compacter show lower average performance (Compacter: BLEU-4 $19.50$, ROUGE-L $36.79$; AdvFusion+Compacter: BLEU-4 $19.52$, ROUGE-L $36.82$), while AdapterFusion+Compacter sits between these groups (BLEU-4 $19.48$, ROUGE-L $37.61$).

Under the hood, AdapterFusion is the best or tied-best configuration on every evaluated model (DeepSeek-Coder 1.3B, Qwen2.5-Coder 1.5B, Qwen2.5-Coder 3B, CodeLlama-7B), and it produces the largest gains on larger models in both BLEU-4 and ROUGE-L. LoRA and TaskAdapter form the next tier of strong baselines, delivering robust performance with relatively low parameter overhead. AdvFusion frequently improves over other baselines, including LoRA on a per-model basis, but does not surpass AdapterFusion on any of the models in our CMG experiments.

\paragraph{\textbf{AdvFusion's performance on Code-LLMs.}}
AdvFusion does not consistently improve CMG performance relative to AdapterFusion. AdvFusion trails AdapterFusion by roughly $0.9$ and $2.8$ percentage points in BLEU-4 and ROUGE-L, respectively. Compared to strong single-method baselines, AdvFusion is competitive with LoRA and TaskAdapter. The per-model inspection shows that AdvFusion is closest to the best performer on smaller models (e.g., DeepSeek-Coder 1.3B, where AdvFusion reaches BLEU-4 $22.53$ and ROUGE-L $40.18$), but it still fails to outpace AdapterFusion and often loses small amounts of performance to LoRA/TaskAdapter. In short, AdvFusion does not provide uniform gains for CMG and is outperformed by AdapterFusion.

\paragraph{\textbf{Impact of replacing Bottleneck adapters with Compacter in the AdvFusion architecture.}}
Replacing Bottleneck adapters with Compacter \emph{did not} improve AdvFusion’s performance. 
Note that this result is expected as the Compacter has the lowest scores for CMG compared to other PEFT approaches.
Although adding AdvFusion to Compacter improves Compacter performance on average, AdvFusion+Compacter achieves an average lower BLEU-4 ($19.52$) and ROUGE-L ($36.82$) than AdvFusion ($21.75$ and $39.76$, respectively). Concretely, AdvFusion+Compacter shows an average decline of roughly $2.2$ and $2.9$ percentage points on BLEU-4 and ROUGE-L, respectively, compared to AdvFusion. Although it is worth noting that the replacement yields a small BLEU-4 improvement only on Qwen2.5Coder 3B, this is an isolated case; on the other three models, the Compacter substitution was seen to have marked drops (e.g., DeepSeekCoder 1.3B: AdvFusion $22.53$ vs. AdvFusion+Compacter $40.18$). 
A similar trend is observed for AdapterFusion+Compacter, where the performance is dropped compared to AdapterFusion. 
In general, replacing bottleneck adapters with Compacter in AdvFusion does not lead to performance improvements for CMG.

\begin{table}[ht]
  \centering
  \caption{BLEU-4 and ROUGE-L results for CMG with different configurations on \textbf{Scala} as the target language.}
  \label{tab:cmg_bleu_rouge_scala}
  \begin{tabular}{llcc}
    \toprule
    \textbf{Model} & \textbf{Configuration} & \textbf{BLEU-4} & \textbf{ROUGE-L} \\
    \midrule
    \multirow{7}{*}{DeepSeek-Coder 1.3B}  
      & AdvFusion+Compacter & 15.31 & 30.38 \\
      & AdvFusion & 22.53 & 40.18 \\
      & AdapterFusion+Compacter & 17.85 & 34.29 \\
      & AdapterFusion & \textbf{22.91} & \textbf{42.80} \\
      & Compacter & 15.30 & 30.64 \\
      & TaskAdapter & \underline{22.70} & \underline{40.51} \\
      & LoRA & 22.07 & 39.46 \\
    \midrule
    \multirow{7}{*}{Qwen2.5-Coder 1.5B}  
      & AdvFusion+Compacter & 20.42 & 38.07 \\
      & AdvFusion & \underline{21.56} & 38.95 \\
      & AdapterFusion+Compacter & 19.33 & 37.26 \\
      & AdapterFusion & \textbf{21.78} & \textbf{41.18} \\
      & Compacter & 20.09 & 37.80 \\
      & TaskAdapter & 21.31 & 38.20 \\
      & LoRA & 21.53 & \underline{39.58} \\
    \midrule
    \multirow{7}{*}{Qwen2.5-Coder 3B} 
      & AdvFusion+Compacter & 21.85 & 39.97 \\
      & AdvFusion & 20.54 & 39.34 \\
      & AdapterFusion+Compacter & 20.75 & 39.25 \\
      & AdapterFusion & \textbf{22.71} & \textbf{43.02} \\
      & Compacter & 21.82 & 39.79 \\
      & TaskAdapter & 21.15 & 40.08 \\
      & LoRA & \underline{22.37} & \underline{41.08} \\
    \midrule
    \multirow{7}{*}{CodeLlama-7B}  
      & AdvFusion+Compacter & 20.50 & 38.85 \\
      & AdvFusion & 22.35 & 40.59 \\
      & AdapterFusion+Compacter & 19.99 & 39.64 \\
      & AdapterFusion & \textbf{23.12} & \textbf{43.41} \\
      & Compacter & 20.78 & 38.94 \\
      & TaskAdapter & \underline{22.89} & \underline{41.41} \\
      & LoRA & 21.84 & 41.14 \\
      \midrule
      \midrule
      \multirow{7}{*}{Average} 
      & AdvFusion+Compacter & 19.52 & 36.82 \\
      & AdvFusion & 21.75 & 39.76 \\
      & AdapterFusion+Compacter & 19.48 & 37.61 \\
      & AdapterFusion & \textbf{22.63} & \textbf{42.60} \\
      & Compacter & 19.50 & 36.79 \\
      & TaskAdapter & \underline{22.01} & 40.05 \\
      & LoRA & 21.95 & \underline{40.31} \\
    \bottomrule
  \end{tabular}
\end{table}

\begin{mdframed} [backgroundcolor=gray!20, linewidth=1pt]
\textbf{In commit message generation, AdapterFusion performed best, with LoRA and TaskAdapter showing competitive performance. Replacing Bottleneck adapters with Compacter in AdvFusion did not enhance AdvFusion's performance.}
\end{mdframed}

\subsection{Code Generation}
\label{sec:results-code-generation}

Table~\ref{tab:code_generation} presents the results of code generation tasks across different models and configurations \footnote{We evaluated performance using pass@k, but the results were all 0, probably due to the complexity of this new public dataset. }. When comparing AdvFusion with AdapterFusion, we observed that AdvFusion achieves improved BLEU and ROUGE scores across all code-LLMs. AdvFusion achieves a higher average performance compared to AdapterFusion, with the BLEU score being 35.1\% higher and the ROUGE score 32.1\% higher. However, when comparing AdvFusion with other PEFT methods, LoRA and TaskAdapter, we find that the other PEFT methods outperform AdvFusion in many cases. On average, TaskAdapter achieves the highest BLEU (21.35) and ROUGE (26.81). Overall, AdvFusion outperforms AdapterFusion, with TaskAdapter generally achieving the best performance.

\paragraph{\textbf{AdvFusion's performance on Code-LLMs.}}
For code generation, Advfusion is consistently more effective than AdapterFusion for code generation across all Code-LLMs. 
For CodeLlama 7B, AdvFusion’s performance decreases compared to its own performance on other models, suggesting that CodeLlama's architecture may negatively affect its effectiveness on code generation. 
However, comparing the two Qwen2.5-Coder variants, AdvFusion was observed to have an increase in performance as the model size increases. 

In most cases, LoRA, Compacter and TaskAdapter outperform AdvFusion in both BLEU and ROUGE metrics. However, for LoRA and TaskAdapter, larger models achieve better performance. The largest improvements were observed for CodeLlama 7B, where LoRA improved the BLEU and ROUGE scores by 47\% and 41\%, respectively, while TaskAdapter achieves the highest overall performance in BLEU at 25.06 and ROUGE at 31.24, corresponding to approximately 89\% and 61\% improvements over AdvFusion. For the other models, the trend is similar but less pronounced.

\paragraph{\textbf{Impact of replacing Bottleneck adapters with Compacter in the AdvFusion architecture.}}
Replacing bottleneck adapters with Compacter in the AdvFusion architecture has a different impact across Code-LLMs. For Qwen2.5-Coder 1.5B, AdvFusion+Compacter shows an improvement in BLEU from 14.6 to 16.8 (an increase of 15.1\%), and ROUGE from 21.6 to 29.6 (an increase of 36.1\%). However, for other Code-LLMs, such as DeepSeek-Coder 1.3B, Qwen2.5-Coder 3B, and CodeLlama 7B, replacing bottleneck adapters with Compacter does not lead to consistent improvements, with decreased performance in BLEU and ROUGE (except for two cases of an increase in ROUGE).

\color{black}

\begin{table}[ht]
  \centering
  \caption{BLEU-4 and ROUGE-L results for code generation tasks with different configurations on PHP as the target language.}
  \label{tab:code_generation}
  \begin{tabular}{llcc}
    \toprule
    \textbf{Model} & \textbf{Configuration} & \textbf{BLEU-4} & \textbf{ROUGE-L} \\

    \midrule
    \multirow{6}{*}{DeepSeek-Coder 1.3B} 
      & AdvFusion+Compacter & 10.66 & 15.34\\
      & AdvFusion & 14.52 & 19.53 \\
      & AdapterFusion+Compacter & 10.16 & 14.32 \\
      & AdapterFusion & 9.40 & 13.44 \\
      & Compacter & 13.94 & 18.38 \\
      & TaskAdapter & \textbf{18.07} & \textbf{23.56}  \\
      & LoRA & \underline{16.25} & \underline{21.76} \\
    \midrule
    \multirow{6}{*}{Qwen2.5-Coder 1.5B} 
      & AdvFusion+Compacter & 16.83 & \textbf{29.62} \\
      & AdvFusion & 14.63 & 21.56 \\
      & AdapterFusion+Compacter & 15.86 & 22.73 \\
      & AdapterFusion & 10.99 & 15.13 \\
      & Compacter & \underline{21.51} & 25.98 \\
      & TaskAdapter & \textbf{24.32} & \underline{28.73} \\
      & LoRA & 16.19 & 22.45 \\
    \midrule
    \multirow{6}{*}{Qwen2.5-Coder 3B} 
      & AdvFusion+Compacter & 15.90 & 26.17 \\
      & AdvFusion & 18.64 & 24.05 \\
      & AdapterFusion+Compacter & 12.15 & 20.09 \\
      & AdapterFusion & 14.35 & 19.63 \\
      & Compacter & \underline{20.89} & \underline{27.33} \\
      & TaskAdapter & 17.97 & 23.71 \\
      & LoRA & \textbf{23.68} & \textbf{30.24} \\
    \midrule
    \multirow{6}{*}{CodeLlama 7B} 
      & AdvFusion+Compacter & 11.49 & 18.17 \\
      & AdvFusion & 13.28 & 19.38 \\
      & AdapterFusion+Compacter & 11.21 & 7.89 \\
      & AdapterFusion & 10.45 & 15.82 \\
      & Compacter & 17.87 & 24.24 \\
      & TaskAdapter & \textbf{25.06} & \textbf{31.24} \\
      & LoRA & \underline{19.63} & \underline{27.34} \\
    \midrule
    \midrule
    \multirow{7}{*}{Average} 
      & AdvFusion+Compacter & 13.72 & 22.32 \\
      & AdvFusion & 15.27 & 21.13 \\
      & AdapterFusion+Compacter & 12.35 & 16.26 \\
      & AdapterFusion & 11.30 & 16.00 \\
      & Compacter & 18.55 & 23.98 \\
      & TaskAdapter & \textbf{21.35} & \textbf{26.81} \\
      & LoRA & \underline{18.94} & \underline{25.45} \\
    \bottomrule
  \end{tabular}
\end{table}

\begin{mdframed} [backgroundcolor=gray!20, linewidth=1pt]
\textbf{In code generation, AdvFusion outperformed AdapterFusion, but overall, TaskAdapter achieved the best performance. The impact of replacing Bottleneck adapters with Compacter in AdvFusion varies across Code-LLMs. }
\end{mdframed}

\subsection{Code Translation}

\begin{table}[!htp]
\centering
\caption{Average BLEU-4 and ROUGE-L scores measured on the test split and Pass@1 and Pass@10 scores measured on PolyHumanEval for code translation across all target languages.}
\label{tab:ct-results}
\resizebox{\textwidth}{!}{ 
\begin{tabular}{llcccc}
\toprule
\textbf{Model} &\textbf{Configuration} &\textbf{BLEU-4} &\textbf{ROUGE-L} &\textbf{Pass@1} &\textbf{Pass@10} \\
    \midrule
    \multirow{7}{*}{DeepSeek-Coder 1.3B} 
    & AdvFusion+Compacter & 10.10 & 21.95 & \underline{27.38} & 47.05 \\
    &AdvFusion &10.03 &22.15 &24.08 &39.65 \\
    &AdapterFusion+Compacter &\underline{10.65} &\underline{22.30} &27.35 &\textbf{47.75} \\
    &AdapterFusion &10.30 &\underline{22.30} &23.08 &39.18 \\
    &Compacter &10.58 &22.28 &\textbf{27.50} &\underline{47.10} \\
    &TaskAdapter &9.98 &21.93 &23.20 &39.30 \\
    &LoRA &\textbf{10.75} &\textbf{22.90} &23.55 &42.25 \\
    \midrule 
    \multirow{7}{*}{Qwen2.5-Coder 1.5B} 
    &AdvFusion+Compacter &9.70 &19.68 &27.33 &45.95 \\
    &AdvFusion &9.78 &20.03 &25.23 &44.48 \\
    &AdapterFusion+Compacter &9.28 &17.90 &24.03 &\underline{47.25} \\
    &AdapterFusion &\underline{10.90} &\underline{21.25} &24.83 &43.40 \\
    &Compacter &9.83 &19.73 &\underline{27.58} &45.50 \\
    &TaskAdapter &8.88 &19.48 &24.98 &42.28 \\
    &LoRA &\textbf{12.78} &\textbf{24.05} &\textbf{30.68} &\textbf{51.00} \\
    \midrule
    \multirow{7}{*}{Qwen2.5-Coder 3B} 
    &AdvFusion+Compacter &11.25 &21.75 &32.70 &51.18 \\
    &AdvFusion &11.00 &21.20 &31.90 &52.95 \\
    &AdapterFusion+Compacter &10.40 &19.90 &20.93 &43.20 \\
    &AdapterFusion &\underline{11.78} &\underline{22.40} &\underline{35.15} &\underline{56.18} \\
    &Compacter &11.00 &21.58 &33.23 &51.20 \\
    &TaskAdapter &9.53 &20.43 &29.58 &50.43 \\
    &LoRA &\textbf{13.78} &\textbf{25.35} &\textbf{37.53} &\textbf{60.65} \\
    \midrule
    \multirow{7}{*}{CodeLlama 7B}
    &AdvFusion+Compacter &\underline{11.50} &22.95 &\textbf{32.50} &\textbf{55.23} \\
    &AdvFusion &7.78 &17.83 &12.30 &23.13 \\
    &AdapterFusion+Compacter &10.83 &22.00 &26.80 &50.88 \\
    &AdapterFusion &7.10 &17.58 &14.10 &23.68 \\
    &Compacter &11.18 &\underline{23.00} &\underline{30.88} &\underline{52.95} \\
    &TaskAdapter &8.95 &20.50 &21.90 &38.20 \\
    &LoRA &\textbf{11.60} &\textbf{24.33} &29.13 &51.45 \\
    \midrule
    \midrule
    \multirow{7}{*}{Average} 
    &AdvFusion+Compacter &10.64 &21.58 &\underline{29.98} &\underline{49.85} \\
    &AdvFusion &9.64 &20.30 &23.38 &40.05 \\
    &AdapterFusion+Compacter &10.29 &20.53 &24.78 &47.27 \\
    &AdapterFusion &10.02 &20.88 &24.29 &40.61 \\
    &Compacter &\underline{10.64} &\underline{21.64} &29.79 &49.19 \\
    &TaskAdapter &9.33 &20.58 &24.91 &42.55 \\
    &LoRA &\textbf{12.23} &\textbf{24.16} &\textbf{30.22} &\textbf{51.34} \\
\bottomrule
\end{tabular}
}
\end{table}

Table \ref{tab:ct-results} summarizes the results of the code translation task on Code-LLMs. The main functionality correctness metrics on the PolyHumanEval benchmark (i.e., Pass@1 and Pass@10) show that, overall, LoRA achieves the strongest average performance across all models, followed closely by AdvFusion+Compacter and Compacter. LoRA leads with the highest average Pass@1 ($30.22$) and Pass@10 ($51.34$), while AdvFusion+Compacter ($29.98$ and $49.85$) and Compacter ($29.79$ and $49.19$) deliver comparable functionality correctness despite their smaller parameter overhead. As the model size decreases, the performance gap among the fine-tuning methods narrows, suggesting that for smaller models, the PEFT architecture plays a less influential role.

For the textual similarity metrics (BLEU-4 and ROUGE-L), LoRA again shows clear dominance, achieving the highest average BLEU-4 ($12.23$) and ROUGE-L ($24.16$) scores. Both AdvFusion+Compacter and Compacter follow closely, with BLEU-4 and ROUGE-L averages at $10.64$ for both methods, and $21.58$ and $21.64$ for AdvFusion+Compacter and Compacter, respectively. This consistency indicates that Compacter-based approaches are highly competitive not only in functional correctness but also in textual alignment with the reference code, while AdapterFusion-based methods show comparatively lower performance across all metrics for code translation.

\paragraph{\textbf{AdvFusion's performance on Code-LLMs.}}
For code translation, AdvFusion exhibits substantially lower performance compared to other fine-tuning methods. Averaged across all models, AdvFusion trails LoRA by nearly $22.6\%$ in Pass@1 and over $21.9\%$ in Pass@10, reflecting consistent underperformance. On the smallest model, DeepSeek-Coder 1.3B, AdvFusion surpasses LoRA and achieves moderate results (Pass@1 of $24.08$, Pass@10 of $39.65$) but remains inferior to its Compacter-based variants. In contrast, on the largest model, CodeLlama 7B, AdvFusion significantly falls behind, dropping to $12.30$ Pass@1 and $23.13$ Pass@10, far below LoRA and AdvFusion+Compacter.

When compared directly to its baseline, AdapterFusion, AdvFusion performs comparably on smaller models, showing a balanced trade-off between functional and textual metrics. On DeepSeek-Coder 1.3B and Qwen2.5-Coder 1.5B, AdvFusion achieves slightly higher Pass@1 and Pass@10 scores, while AdapterFusion attains marginally better BLEU-4 and ROUGE-L performance. This indicates that, for lower-capacity models, both methods fine-tune representations similarly. However, as the model size increases, on Qwen2.5-Coder 3B and CodeLlama 7B, the performance gap gradually increases in favour of AdapterFusion, suggesting that AdvFusion’s effectiveness for code translation diminishes as model scale increases.

Overall, AdvFusion remains less effective for code translation on Code-LLMs, with performance inversely correlated with model scale. Moreover, AdvFusion exhibits subpar average performance compared to AdapterFusion on Code-LLMs.

\paragraph{\textbf{Impact of replacing Bottleneck adapters with Compacter in the AdvFusion architecture.}}
Replacing Bottleneck adapters with Compacter modules in the AdvFusion architecture yields a substantial and consistent performance improvement across all models. The AdvFusion+Compacter variant not only bridges most of the gap between AdvFusion and the stronger methods but often surpasses baseline configurations in both functionality and textual metrics. On average, it improves over AdvFusion by $28.2\%$ on Pass@1, $24.4\%$ on Pass@10, and $10.3\%$ on BLEU-4 points. These gains are most pronounced on CodeLlama 7B, where AdvFusion+Compacter boosted Pass@1 from 12.30 to 32.50 and Pass@10 from 23.13 to 55.23.

Overall, substituting bottleneck adapters with Compacter modules significantly enhances AdvFusion’s performance, producing models that surpass AdvFusion and AdapterFusion, and perform competitively with the most capable fine-tuning method, LoRA.

\begin{mdframed} [backgroundcolor=gray!20, linewidth=1pt]
\textbf{In code translation, AdvFusion performed worse than AdapterFusion overall, while LoRA achieved the best performance. Replacing Bottleneck adapters with Compacter in AdvFusion consistently improves its performance across Code-LLMs.}
\end{mdframed}

\section{Discussion} \label{section:discussion}

In this section, we present the results of additional experiments, which provide insights about the AdvFusion versus other PEFT methods, including language-specific performance trends and the contribution of programming languages for a target programming language.
Finally, we offer practical guidance on AdvFusion, when to use this PEFT method and when not to.

\subsection{Language Family Alignment Improves CMG Performance}
We observed during CMG experiments, when replacing Ruby and Coffeescript with C and Java, which are in similar families with the rest of the programming languages in our dataset (Rust, Swift, and Scala), there is a performance improvement in BLEU-4 and ROUGE-L. For instance, DeepSeek-Coder 1.3B AdvFusion, BLEU-4 and ROUGE-L scores for \textbf{Swift} increase from \textbf{14.40} and \textbf{29.30} to \textbf{15.57} and \textbf{33.43}, respectively. This indicates that programming languages that share structural or family-level similarities (e.g., C--C++ or Java--Kotlin) promote better cross-language generalization and downstream effectiveness. This supports the hypothesis that syntactic and semantic alignment across source programming languages enhances the fusion process.

\subsection{Comparing Code-Only and Commit-Message Training for CMG}
For CMG, we also trained Compacter on code-only CommitPack subsets (omitting commit messages, statistics in Table \ref{tab:commitpack-stats}) and then ran AdapterFusion/AdvFusion on the CMG dataset. This variant is motivated by the hypothesis that low-level code-only signals could bootstrap adapter representations before teaching them commit-message semantics \cite{muennighoff2023octopack}. We observed that this pipeline produced inferior results compared to training Compacter on commit-message pairs. For instance, with \textbf{Swift} as the target language, Qwen2.5Coder 1.5B AdapterFusion+Compacter achieved \textbf{16.72} BLEU-4 and \textbf{40.38} ROUGE-L when Compacter is also trained on CMG data, surpassing the \textbf{15.50} BLEU-4 and \textbf{37.30} ROUGE-L that were obtained when Compacter is trained on code-only data (using CLM).

\begin{table}[htbp]
\centering
\caption{Dataset statistics for subset used from CommitPack \cite{muennighoff2023octopack} as code-only data.}
\label{tab:commitpack-stats}
\begin{tabular}{lrrrr}
\toprule
\textbf{Language} &\textbf{Train} &\textbf{Test} &\textbf{Total} \\\midrule
Java &48,000 &2,000 &50,000 \\
C &48,000 &2,000 &50,000 \\
Scala &48,000 &2,000 &50,000 \\
Swift &48,000 &2,000 &50,000 \\
Rust &48,000 &2,000 &50,000 \\
\bottomrule
\end{tabular}
\vspace{1ex}
\end{table}

\subsection{Language-specific Performance Trends in PEFT for CMG}

In order to observe the effect of different target programming languages on PEFT methods for CMG, we used ROUGE-L and BLEU-4 scores as well as their averages across methods and programming languages shown in Tables \ref{tab:cmg_all_rouge} and \ref{tab:cmg_all_bleu}. For ROUGE-L (Table~\ref{tab:cmg_all_rouge}), AdapterFusion leads overall (average ROUGE-L of $36.86$), with pronounced strengths on Scala ($42.60$) and Swift ($42.41$). By contrast, AdvFusion has the highest average on Rust ($26.27$) while LoRA on CodeLLama yields $29.33$ on Rust.

BLEU-4 (Table~\ref{tab:cmg_all_bleu}) provides a complementary view: AdvFusion achieves the highest BLEU-4 average on Rust ($19.04$) and AdapterFusion leads BLEU specifically on Scala ($22.63$). The variation in BLEU-4 scores is smaller than the ROUGE-L gap between the highest and lowest values (ROUGE-L $\approx 21.6\%$; BLEU-4 $\approx 12.6\%$ relative to the smallest average score), indicating that the choice of PEFT influences long-sequence semantic fidelity more strongly than short n-gram precision. TaskAdapter peaks on BLEU-4 overall.

\textbf{Practical Takeaway:} In CMG, \emph{AdapterFusion} is preferred when the target programming language is Scala or Swift. However, for Rust, \emph{AdvFusion} or \emph{TaskAdapter} should be considered. We also note that with larger model size, specifically, AdapterFusion's ROUGE-L improves from $36.20$ to $38.41$ from DeepSeek-Coder 1.3B to CodeLlama 7B, respectively. Thus, we advocate considering both the target programming language and the base model for choosing a PEFT method in CMG.

\begin{table}[!htp]
\centering
\caption{ROUGE-L scores of Code-LLMs on CMG to target languages Rust, Scala, Swift.}
\resizebox{\textwidth}{!}{
\begin{tabular}{llcccc}
\toprule
\textbf{Model} & \textbf{Configuration} & \textbf{Rust} & \textbf{Scala} & \textbf{Swift} & \textbf{Average} \\
\midrule
\multirow{7}{*}{DeepSeek-Coder 1.3B} 
 & AdvFusion+Compacter & 19.72 & 30.38 & 29.19 & 26.43 \\
 & AdvFusion & \textbf{24.54} & 40.18 & 33.43 & 32.71 \\
 & AdapterFusion+Compacter & 20.17 & 34.29 & 30.24 & 28.24 \\
 & AdapterFusion & \underline{24.18} & \textbf{42.80} & \textbf{41.60} & \textbf{36.20} \\
 & Compacter & 19.55 & 30.64 & 29.15 & 26.45 \\
 & TaskAdapter & 23.28 & \underline{40.51} & 33.68 & 32.49 \\
 & LoRA & 23.65 & 39.46 & \underline{37.06} & \underline{33.39} \\
\midrule
\multirow{7}{*}{Qwen2.5-Coder 1.5B} 
 & AdvFusion+Compacter & 24.91 & 38.07 & 32.79 & 31.92 \\
 & AdvFusion & \textbf{25.34} & 38.95 & 33.71 & 32.67 \\
 & AdapterFusion+Compacter & 22.35 & 37.26 & \underline{40.38} & \underline{33.33} \\
 & AdapterFusion & 24.13 & \textbf{41.18} & \textbf{42.12} & \textbf{35.81} \\
 & Compacter & 24.48 & 37.80 & 33.85 & 32.04 \\
 & TaskAdapter & \underline{25.03} & 38.20 & 30.55 & 31.26 \\
 & LoRA & 24.48 & \underline{39.58} & 30.35 & 31.47 \\
\midrule
\multirow{7}{*}{Qwen2.5-Coder 3B} 
 & AdvFusion+Compacter & 23.26 & 39.97 & 26.86 & 30.03 \\
 & AdvFusion & 26.29 & 39.34 & 35.22 & 33.62 \\
 & AdapterFusion+Compacter & 22.71 & 39.25 & \underline{37.52} & 33.16 \\
 & AdapterFusion & \textbf{26.73} & \textbf{43.02} & \textbf{41.30} & \textbf{37.02} \\
 & Compacter & 23.56 & 39.79 & 29.86 & 31.07 \\
 & TaskAdapter & \underline{26.64} & 40.08 & 35.44 & \underline{34.05} \\
 & LoRA & 26.18 & \underline{41.08} & 34.04 & 33.76 \\
\midrule
\multirow{7}{*}{CodeLlama 7B} 
 & AdvFusion+Compacter & 25.39 & 38.85 & 34.26 & 32.83 \\
 & AdvFusion & \underline{28.92} & 40.59 & 33.20 & 34.24 \\
 & AdapterFusion+Compacter & 24.53 & 39.64 & \underline{36.87} & 33.68 \\
 & AdapterFusion & 27.21 & \textbf{43.41} & \textbf{44.61} & \textbf{38.41} \\
 & Compacter & 25.65 & 38.94 & 32.95 & 32.51 \\
 & TaskAdapter & 28.05 & \underline{41.41} & 33.41 & 34.29 \\
 & LoRA & \textbf{29.33} & 41.14 & 36.82 & \underline{35.76} \\
\midrule
\midrule
\multirow{7}{*}{Average} 
 & AdvFusion+Compacter & 23.32 & 36.82 & 30.78 & 30.31 \\
 & AdvFusion & \textbf{26.27} & 39.76 & 33.89 & 33.31 \\
 & AdapterFusion+Compacter & 22.44 & 37.61 & \underline{36.25} & 32.10 \\
 & AdapterFusion & 25.56 & \textbf{42.60} & \textbf{42.41} & \textbf{36.86} \\
 & Compacter & 23.31 & 36.79 & 31.45 & 30.52 \\
 & TaskAdapter & 25.75 & 40.05 & 33.27 & 33.02 \\
 & LoRA & \underline{25.91} & \underline{40.31} & 34.57 & \underline{33.60} \\
\bottomrule
\end{tabular}
}
\label{tab:cmg_all_rouge}
\end{table}

\begin{table}[!htp]
\centering
\caption{BLEU-4 scores of Code-LLMs on CMG to target languages Rust, Scala, Swift.}
\resizebox{\textwidth}{!}{
\begin{tabular}{llcccc}
\toprule
\textbf{Model} & \textbf{Configuration} & \textbf{Rust} & \textbf{Scala} & \textbf{Swift} & \textbf{Average} \\
\midrule
\multirow{7}{*}{DeepSeek-Coder 1.3B} 
 & AdvFusion+Compacter & 16.24 & 15.31 & 12.03 & 14.53 \\
 & AdvFusion & \textbf{18.34} & 22.53 & \underline{15.57} & \textbf{18.81} \\
 & AdapterFusion+Compacter & 16.21 & 17.85 & 12.83 & 15.63 \\
 & AdapterFusion & 17.80 & \textbf{22.91} & 15.05 & 18.59 \\
 & Compacter & 16.11 & 15.30 & 12.04 & 14.48 \\
 & TaskAdapter & \underline{18.03} & \underline{22.70} & \textbf{15.66} & \underline{18.80} \\
 & LoRA & 17.62 & 22.07 & 14.74 & 18.14 \\
\midrule
\multirow{7}{*}{Qwen2.5-Coder 1.5B} 
 & AdvFusion+Compacter & 18.01 & 20.42 & 15.58 & 18.00 \\
 & AdvFusion & \underline{18.09} & \underline{21.56} & 15.77 & \underline{18.47} \\
 & AdapterFusion+Compacter & 17.67 & 19.33 & \textbf{16.72} & 17.91 \\
 & AdapterFusion & 17.78 & \textbf{21.78} & 15.93 & \textbf{18.50} \\
 & Compacter & 17.96 & 20.09 & \underline{16.07} & 18.04 \\
 & TaskAdapter & \textbf{18.45} & 21.31 & 15.15 & 18.30 \\
 & LoRA & 17.65 & 21.53 & 14.24 & 17.80 \\
\midrule
\multirow{7}{*}{Qwen2.5-Coder 3B} 
 & AdvFusion+Compacter & 17.70 & 21.85 & 13.36 & 17.64 \\
 & AdvFusion & \underline{19.56} & 20.54 & \underline{17.24} & 19.11 \\
 & AdapterFusion+Compacter & 17.47 & 20.75 & 15.71 & 17.98 \\
 & AdapterFusion & 19.14 & \textbf{22.71} & 16.69 & \underline{19.51} \\
 & Compacter & 17.79 & 21.82 & 14.04 & 17.88 \\
 & TaskAdapter & \textbf{19.79} & 21.15 & \textbf{17.70} & \textbf{19.55} \\
 & LoRA & 19.54 & \underline{22.37} & 15.42 & 19.11 \\
\midrule
\multirow{7}{*}{CodeLlama 7B} 
 & AdvFusion+Compacter & 18.58 & 20.50 & 15.84 & 18.31 \\
 & AdvFusion & \textbf{20.18} & 22.35 & 18.42 & 20.32 \\
 & AdapterFusion+Compacter & 18.38 & 19.99 & 15.97 & 18.11 \\
 & AdapterFusion & 18.71 & \textbf{23.12} & \textbf{19.57} & \underline{20.46} \\
 & Compacter & 19.04 & 20.78 & 15.24 & 18.35 \\
 & TaskAdapter & 19.76 & \underline{22.89} & \underline{18.75} & \textbf{20.47} \\
 & LoRA & \underline{19.84} & 21.84 & 17.83 & 19.84 \\
\midrule
\midrule
\multirow{7}{*}{Average} 
 & AdvFusion+Compacter & 17.63 & 19.52 & 14.20 & 17.12 \\
 & AdvFusion & \textbf{19.04} & 21.75 & 16.75 & 19.18 \\
 & AdapterFusion+Compacter & 17.43 & 19.48 & 15.31 & 17.41 \\
 & AdapterFusion & 18.36 & \textbf{22.63} & \underline{16.81} & \underline{19.26} \\
 & Compacter & 17.72 & 19.50 & 14.35 & 17.19 \\
 & TaskAdapter & \underline{19.01} & \underline{22.01} & \textbf{16.81} & \textbf{19.28} \\
 & LoRA & 18.66 & 21.95 & 15.56 & 18.72 \\
\bottomrule
\end{tabular}
}
\label{tab:cmg_all_bleu}
\end{table}

\subsection{Language-specific Performance Trends in PEFT for Code Generation}

We observed that TaskAdapter achieved the best performance for the majority of Code-LLMs, including DeepSeek-Coder, Qwen2.5-Coder 1B, and CodeLlama, except for Qwen2.5-Coder 3B. This suggests that TaskAdapter offers a more effective adaptation mechanism by enabling richer feature transformations while maintaining parameter efficiency in code generation tasks. AdapterFusion and AdvFusion are less suitable for Code-LLMs in code generation. Therefore, we recommend using TaskAdapter for code generation.
Note that this result is based on BLEU and ROUGE scores. But in general, these models are not suitable for generating code that passes test cases (shown by the Pass@K metric).

\subsection{Programming Languages’ Contribution for a Target Programming Language in Code Generation}

\begin{figure}[]
    \centering
    \subfloat[\centering AdapterFusion]{
        \includegraphics[width=1\textwidth]{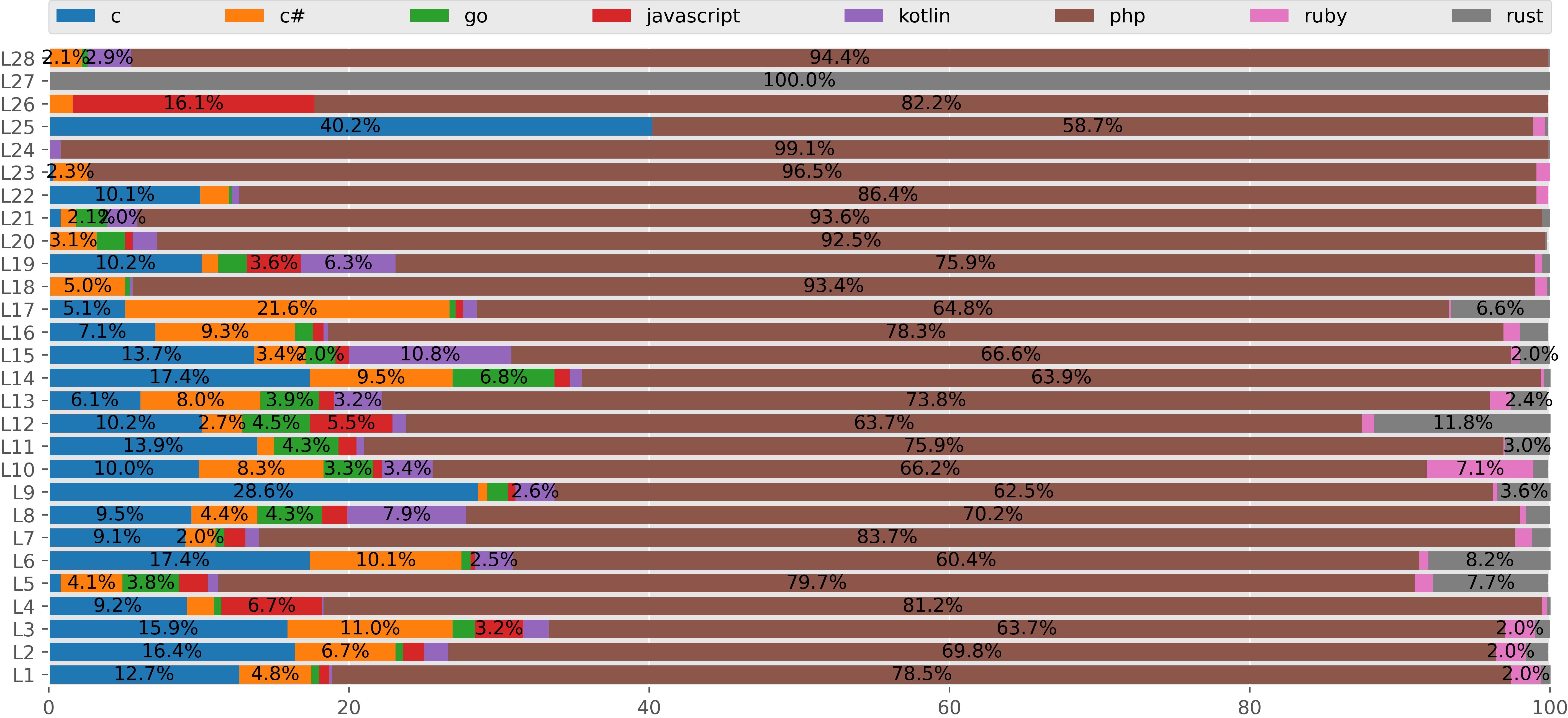}
         \label{fig:qwen_php_fusion}
    }\\
    \subfloat[\centering AdvFusion]{
        \includegraphics[width=1\textwidth]{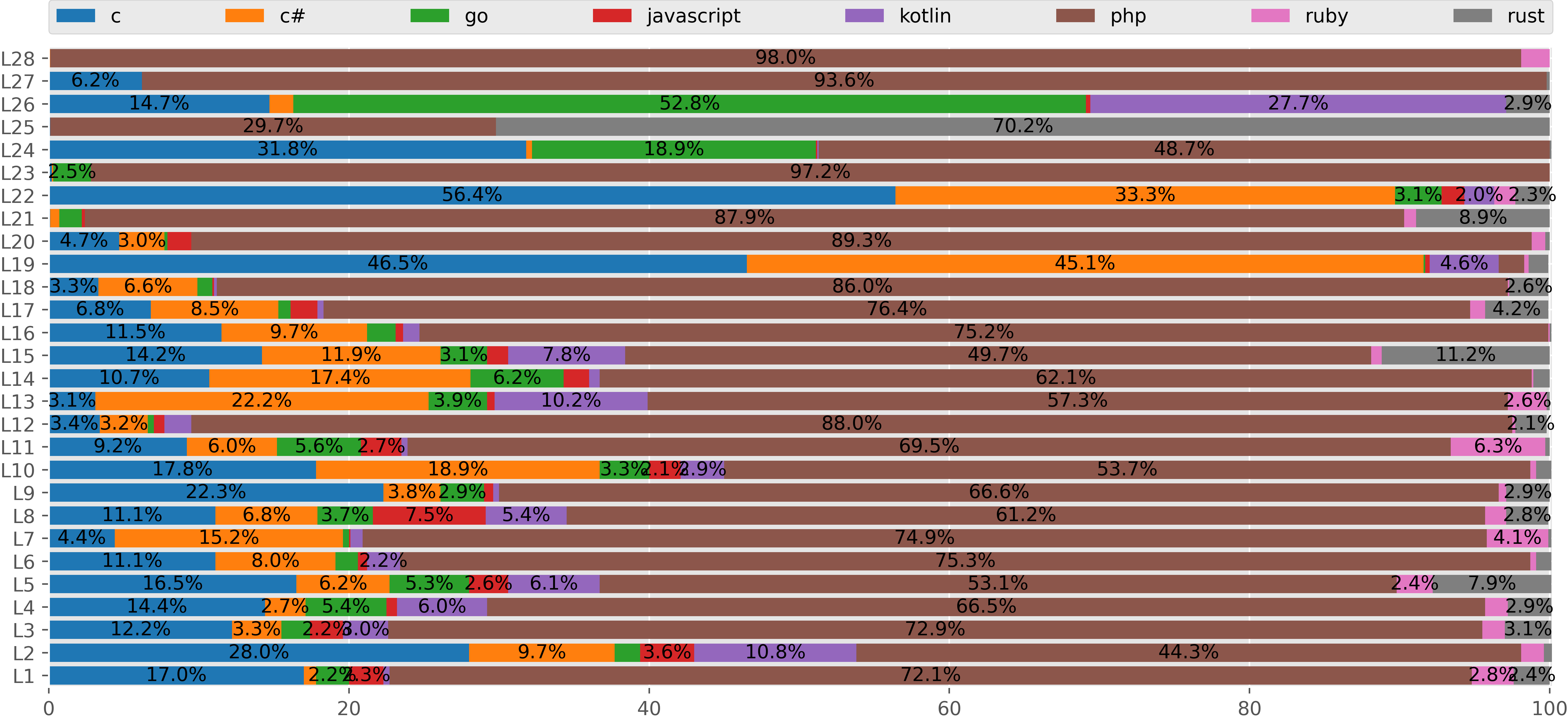}
        \label{fig:qwen_php_advf}
    }
    \caption{The attention contributions from each programming language at each layer in the code generation task for AdapterFusion and AdvFusion using Qwen2.5-Coder 1.5B. The target programming language is PHP.}
    \label{fig:cg-attn-taskadapter}
\end{figure}

Figure \ref{fig:cg-attn-taskadapter} shows the attention contribution of each programming language when AdapterFusion and AdvFusion are used for code generation in PHP. The x-axis indicates the percentage contribution of each programming language, and the y-axis corresponds to the layer number in Qwen2.5-Coder 1.5B. Similar to prior findings, for most layers, a high percentage of attention (more than 80\%) is directed towards the target language PHP (the brown bar) in AdapterFusion, shown in Figure \ref{fig:qwen_php_fusion}. This suggests that AdapterFusion still tends to prioritize the target language, consistent with the behaviour reported in our earlier study \cite{saberi2025advfusion}. Figure \ref{fig:qwen_php_advf} shows the contribution of each programming language in Qwen 2.5 1.5B when PHP is the target language for AdvFusion. AdvFusion allocates more attention to other languages in some higher layers, particularly in layers 19, 22, and 26. For example, in layer 19, the model attends more to C, in layer 22, it also learns more from C, and in layer 26, it learns more from Go. This indicates that, in Code-LLMs, AdvFusion can leverage cross-language knowledge rather than relying exclusively on the target language adapter, as found in our previous work \cite{saberi2025advfusion}.

\begin{figure}[]
    \centering
    \subfloat[\centering AdapterFusion+Compacter]{
        \includegraphics[width=1\textwidth]{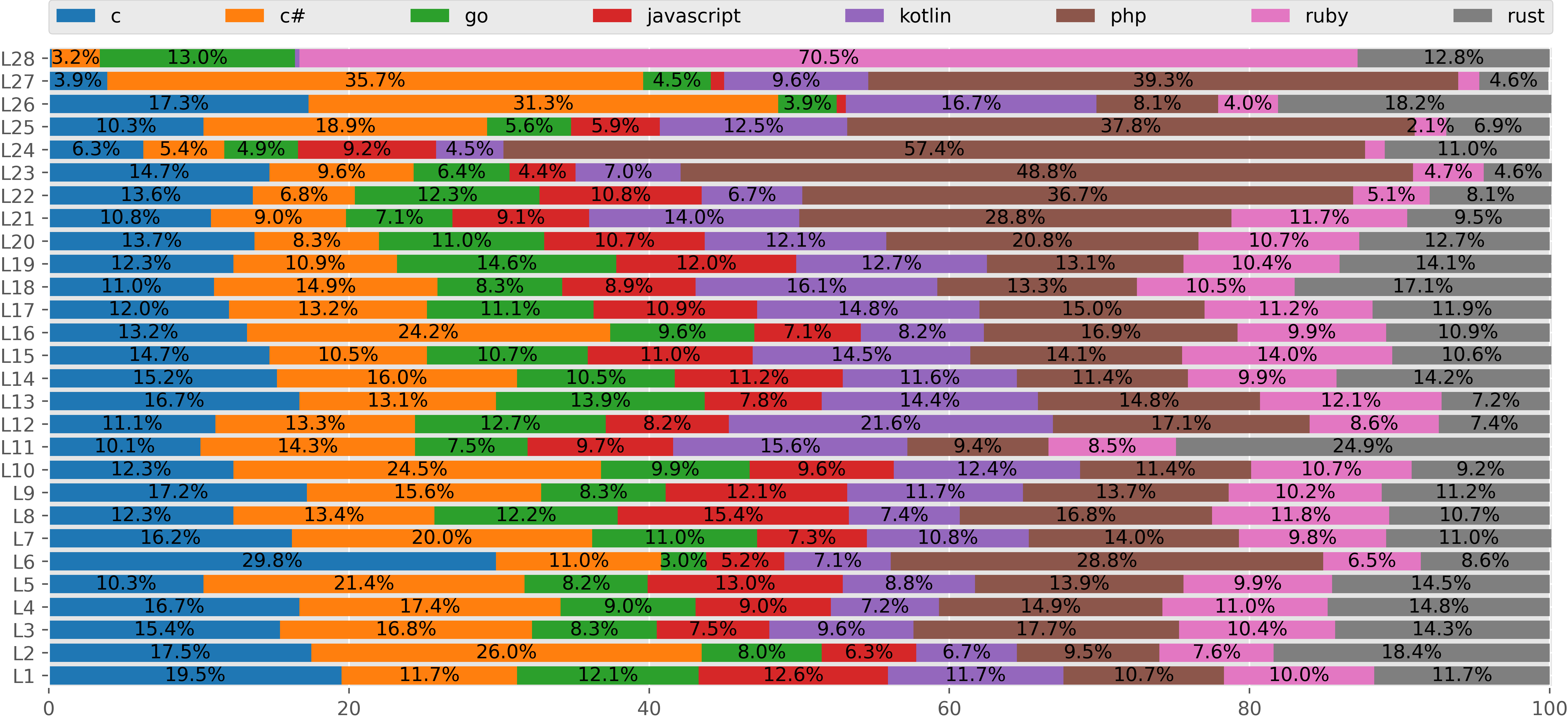}
    }\\
    \subfloat[\centering AdvFusion+Compacter]{
        \includegraphics[width=1\textwidth]{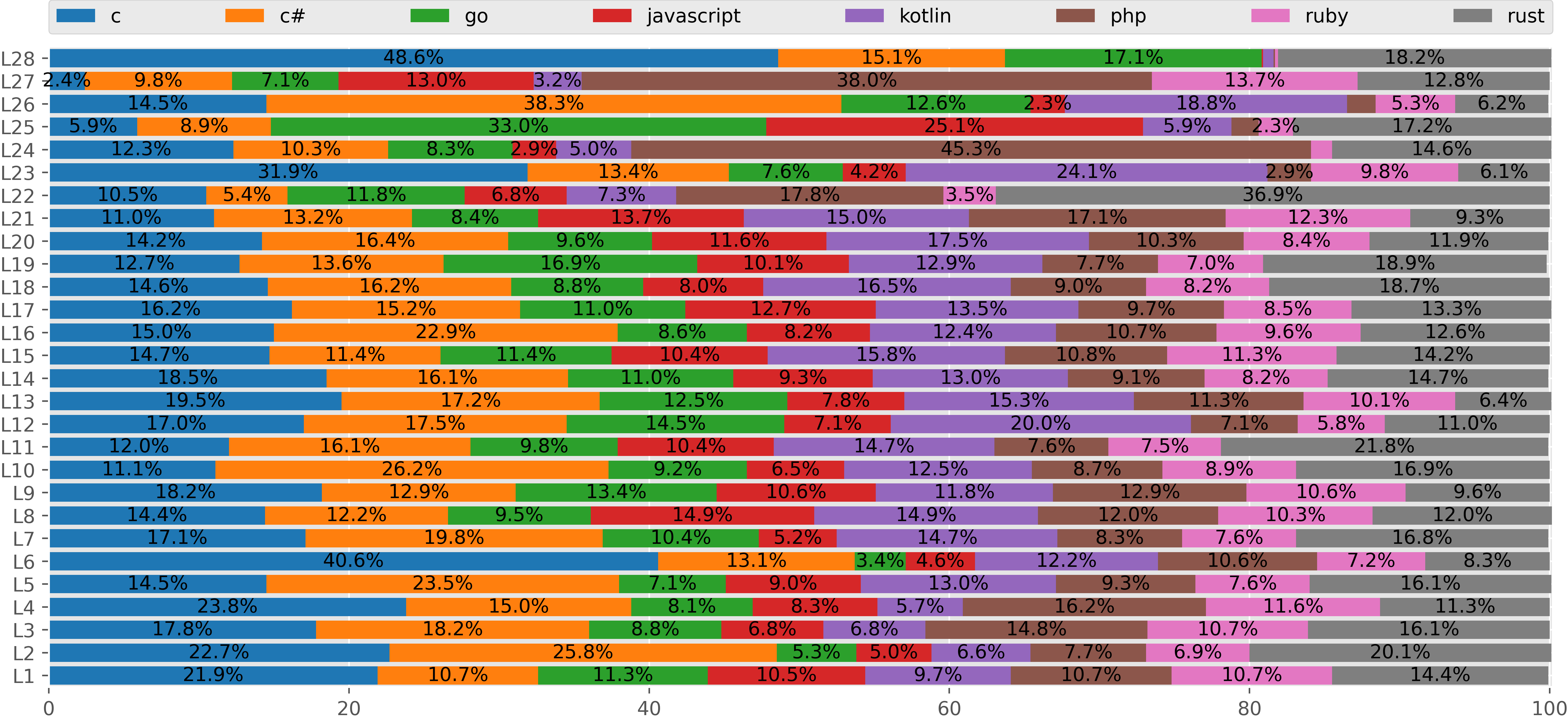}
    }
    \caption{The attention contributions from each programming language at each layer in the code generation task for AdapterFusion+Compacter and AdvFusion+Compacter using Qwen2.5-Coder 1.5B. The target programming language is PHP.}
    \label{fig:cg-attn-compacter}
\end{figure}

Figure \ref{fig:cg-attn-compacter} illustrates the attention contributions of each programming language for AdapterFusion+Compacter and AdvFusion+Compacter in Qwen2.5-Coder 1.5B. The x-axis represents the percentage contribution, and the y-axis shows the layer number in Qwen2.5-Coder 1.5B. Similar to AdapterFusion and AdvFusion, AdvFusion+Compacter pays more attention to other programming languages. However, unlike AdvFusion, AdvFusion+Compacter consistently allocates a larger proportion of attention to other programming languages across all layers.

\subsection{Language-specific Performance Trends in PEFT for Code Translation}

\begin{table}[!htp]
\centering
\caption{Pass@1 scores of Code-LLMs on code translation to target languages Julia, Ruby, Scala and Swift.}
\resizebox{\textwidth}{!}{ 
\begin{tabular}{llcccc}
\toprule
\textbf{Model} &\textbf{Configuration} &\textbf{Julia} &\textbf{Ruby} &\textbf{Scala} &\textbf{Swift} \\
\midrule
\multirow{7}{*}{DeepSeek-Coder 1.3B} &AdvFusion+Compacter &31.80 &\underline{38.70} &15.50 &\underline{23.50} \\
&AdvFusion &\underline{32.60} &34.70 &11.10 &17.90 \\
&AdapterFusion+Compacter &32.00 &\textbf{39.30} &\underline{16.50} &21.60 \\
&AdapterFusion &\textbf{33.00} &33.20 &7.30 &18.80 \\
&Compacter &32.00 &37.70 &\textbf{16.70} &\textbf{23.60} \\
&TaskAdapter &31.30 &32.50 &10.00 &19.00 \\
&LoRA &30.70 &33.00 &15.20 &15.30 \\
\midrule
\multirow{7}{*}{Qwen2.5-Coder 1.5B} &AdvFusion+Compacter &\textbf{36.00} &35.70 &7.60 &30.00 \\
&AdvFusion &31.60 &29.90 &13.90 &25.50 \\
&AdapterFusion+Compacter &24.30 &26.00 &\underline{15.20} &30.60 \\
&AdapterFusion &32.90 &33.60 &6.60 &26.20 \\
&Compacter &\underline{34.70} &\underline{37.20} &7.70 &\underline{30.70} \\
&TaskAdapter &29.20 &\textbf{37.30} &10.00 &23.40 \\
&LoRA &34.00 &31.90 &\textbf{23.80} &\textbf{33.00} \\
\midrule
\multirow{7}{*}{Qwen2.5-Coder 3B} &AdvFusion+Compacter &\textbf{44.10} &42.80 &7.20 &36.70 \\
&AdvFusion &36.20 &33.00 &\underline{26.50} &31.90 \\
&AdapterFusion+Compacter &12.20 &11.40 &22.50 &\underline{37.60} \\
&AdapterFusion &35.40 &\underline{44.90} &\textbf{28.30} &32.00 \\
&Compacter &42.50 &44.60 &7.00 &\textbf{38.80} \\
&TaskAdapter &31.80 &36.00 &26.40 &24.10 \\
&LoRA &\underline{42.80} &\textbf{45.90} &26.30 &35.10 \\
\midrule
\multirow{7}{*}{CodeLlama 7B} &AdvFusion+Compacter &\textbf{41.10} &\underline{41.40} &\textbf{15.90} &\textbf{31.60} \\
&AdvFusion &19.10 &17.10 &2.50 &10.50 \\
&AdapterFusion+Compacter &32.40 &32.30 &12.10 &30.40 \\
&AdapterFusion &23.10 &20.50 &0.80 &12.00 \\
&Compacter &\underline{38.90} &40.30 &12.70 &\textbf{31.60} \\
&TaskAdapter &30.00 &28.60 &9.30 &19.70 \\
&LoRA &37.50 &\textbf{42.00} &\textbf{15.90} &21.10 \\
\midrule
\midrule
\multirow{7}{*}{Average} &AdvFusion+Compacter &\textbf{38.25} &\underline{39.65} &11.55 &\underline{30.45} \\
&AdvFusion &29.88 &28.68 &13.50 &21.45 \\
&AdapterFusion+Compacter &25.23 &27.25 &\underline{16.58} &30.05 \\
&AdapterFusion &31.10 &33.05 &10.75 &22.25 \\
&Compacter &\underline{37.03} &\textbf{39.95} &11.03 &\textbf{31.18} \\
&TaskAdapter &30.58 &33.60 &13.93 &21.55 \\
&LoRA &36.25 &38.20 &\textbf{20.30} &26.13 \\
\bottomrule
\end{tabular}
}
\label{tab:ct_all_pass1}
\end{table}

\begin{table}[!htp]
\centering
\caption{BLEU-4 scores of Code-LLMs on code translation to target languages Julia, Ruby, Scala and Swift.}
\resizebox{\textwidth}{!}{
\begin{tabular}{llcccc}
\toprule
\textbf{Model} &\textbf{Configuration} &\textbf{Julia} &\textbf{Ruby} &\textbf{Scala} &\textbf{Swift} \\
\midrule
\multirow{7}{*}{DeepSeek-Coder 1.3B} &AdvFusion+Compacter &\underline{8.30} &\underline{10.00} &8.80 &\underline{13.30} \\
&AdvFusion &7.10 &8.70 &13.00 &11.30 \\
&AdapterFusion+Compacter &\textbf{8.40} &\textbf{10.20} &11.00 &13.00 \\
&AdapterFusion &7.30 &9.00 &13.80 &11.10 \\
&Compacter &8.20 &9.80 &10.80 &\textbf{13.50} \\
&TaskAdapter &7.00 &8.40 &\underline{14.30} &10.20 \\
&LoRA &6.90 &9.30 &\textbf{15.10} &11.70 \\
\midrule
\multirow{7}{*}{Qwen2.5-Coder 1.5B} &AdvFusion+Compacter &7.20 &5.80 &12.80 &13.00 \\
&AdvFusion &7.40 &5.90 &13.10 &12.70 \\
&AdapterFusion+Compacter &7.40 &6.80 &9.70 &13.20 \\
&AdapterFusion &\underline{7.80} &\underline{7.30} &\underline{15.60} &12.90 \\
&Compacter &7.00 &6.00 &12.50 &\underline{13.80} \\
&TaskAdapter &7.10 &6.30 &12.30 &9.80 \\
&LoRA &\textbf{8.60} &\textbf{9.30} &\textbf{17.90} &\textbf{15.30} \\
\midrule
\multirow{7}{*}{Qwen2.5-Coder 3B} &AdvFusion+Compacter &\underline{8.70} &7.20 &14.30 &14.80 \\
&AdvFusion &7.60 &6.00 &\underline{16.10} &14.30 \\
&AdapterFusion+Compacter &8.40 &6.70 &12.40 &14.10 \\
&AdapterFusion &8.60 &\underline{9.20} &16.00 &13.30 \\
&Compacter &8.40 &7.00 &13.70 &\underline{14.90} \\
&TaskAdapter &7.10 &7.10 &13.40 &10.50 \\
&LoRA &\textbf{9.80} &\textbf{10.30} &\textbf{18.90} &\textbf{16.10} \\
\midrule
\multirow{7}{*}{CodeLlama 7B} &AdvFusion+Compacter &\textbf{9.50} &9.70 &12.40 &\textbf{14.40} \\
&AdvFusion &6.50 &6.10 &10.00 &8.50 \\
&AdapterFusion+Compacter &8.40 &\underline{10.10} &11.40 &\underline{13.40} \\
&AdapterFusion &7.40 &6.60 &6.30 &8.10 \\
&Compacter &\underline{8.60} &9.60 &\underline{13.30} &13.20 \\
&TaskAdapter &7.40 &7.70 &12.30 &8.40 \\
&LoRA &7.50 &\textbf{10.60} &\textbf{15.60} &12.70 \\
\midrule
\midrule
\multirow{7}{*}{Average} &AdvFusion+Compacter &\textbf{8.43} &8.18 &12.08 &\underline{13.88} \\
&AdvFusion &7.15 &6.68 &13.05 &11.70 \\
&AdapterFusion+Compacter &8.15 &\underline{8.45} &11.13 &13.43 \\
&AdapterFusion &7.78 &8.03 &12.93 &11.35 \\
&Compacter &8.05 &8.10 &12.58 &13.85 \\
&TaskAdapter &7.15 &7.38 &\underline{13.08} &9.73 \\
&LoRA &\underline{8.20} &\textbf{9.88} &\textbf{16.88} &\textbf{13.95} \\
\bottomrule
\end{tabular}
}
\label{tab:ct_all_bleu}
\end{table}

Examining the performance of PEFT methods across different target languages for code translation reveals interesting language-specific patterns. For functionality correctness (Pass@1) presented in Table~\ref{tab:ct_all_pass1}, Compacter and AdvFusion+Compacter generally lead on Julia, Ruby, and Swift, often outperforming LoRA and other configurations. In contrast, Scala exhibits a distinct trend: LoRA and AdapterFusion+Compacter achieve the highest Pass@1 scores, while AdvFusion+Compacter performs relatively poorly compared to its results on the other programming languages. This suggests that certain PEFT methods may better capture the structural characteristics of specific programming languages.

For textual similarity (BLEU-4) presented in Table~\ref{tab:ct_all_bleu}, LoRA consistently achieves the highest scores across all programming languages, indicating a robust alignment with the reference code regardless of the target programming language. Other methods, such as AdvFusion+Compacter and AdvFusion, perform similarly to one another, with only minor variations between programming languages. Notably, the differences in BLEU-4 are less pronounced than in functionality correctness, suggesting that textual alignment is less sensitive to the choice of PEFT method for a given programming language.

\textbf{Practical Takeaway:} Overall, programming language-specific analysis highlights that performance trends among fine-tuning methods can vary depending on the target programming language in code translation, particularly in functionality metrics, emphasizing the importance of considering target programming language characteristics when selecting a PEFT strategy.

\subsection{Programming Languages’ Contribution for a Target Programming Language in Code Translation}

\begin{figure}
    \centering
    \subfloat[\centering AdapterFusion]{{\includegraphics[width=\textwidth]{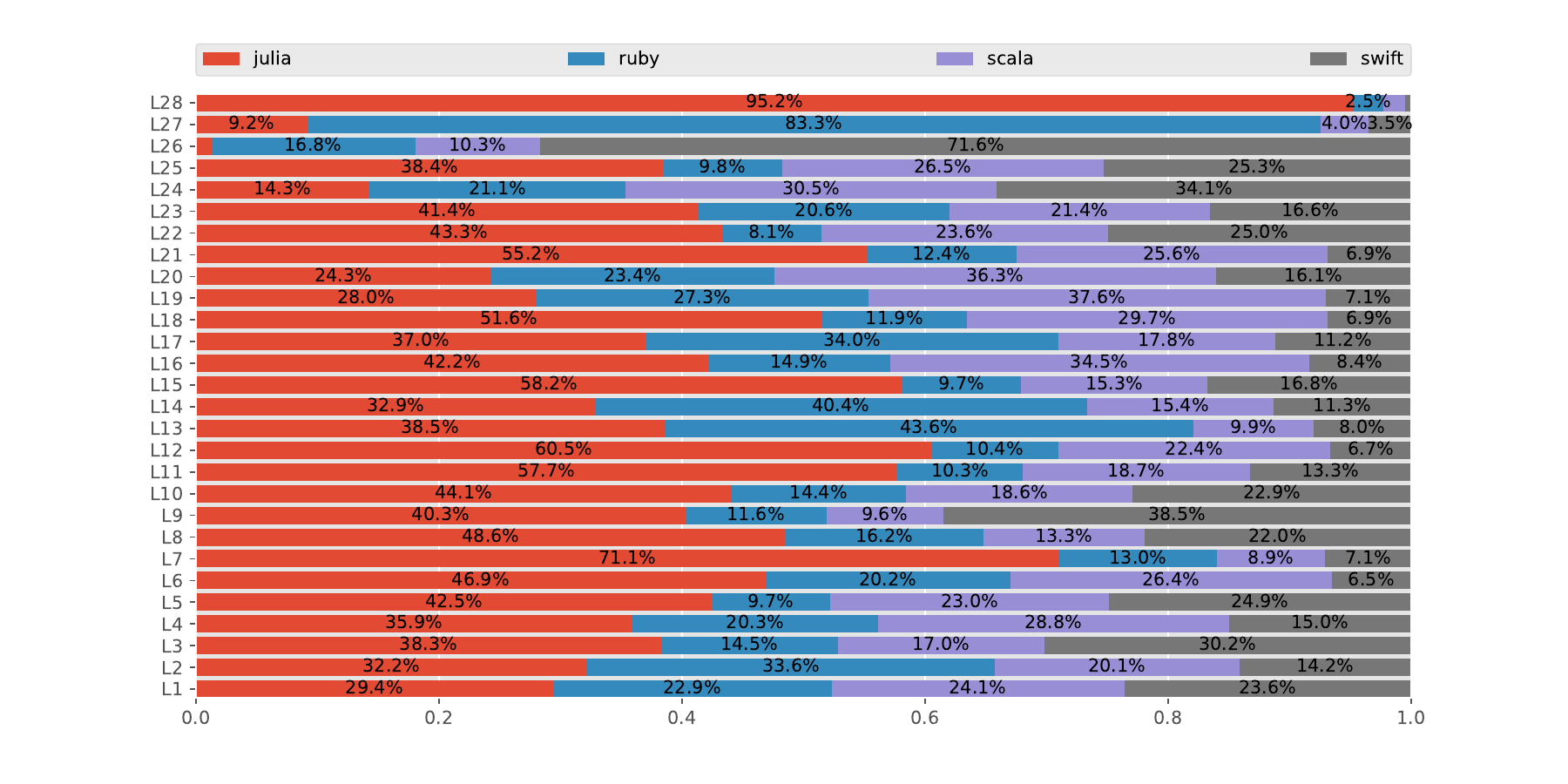}}}\\
    \subfloat[\centering AdvFusion]{{\includegraphics[width=\textwidth]{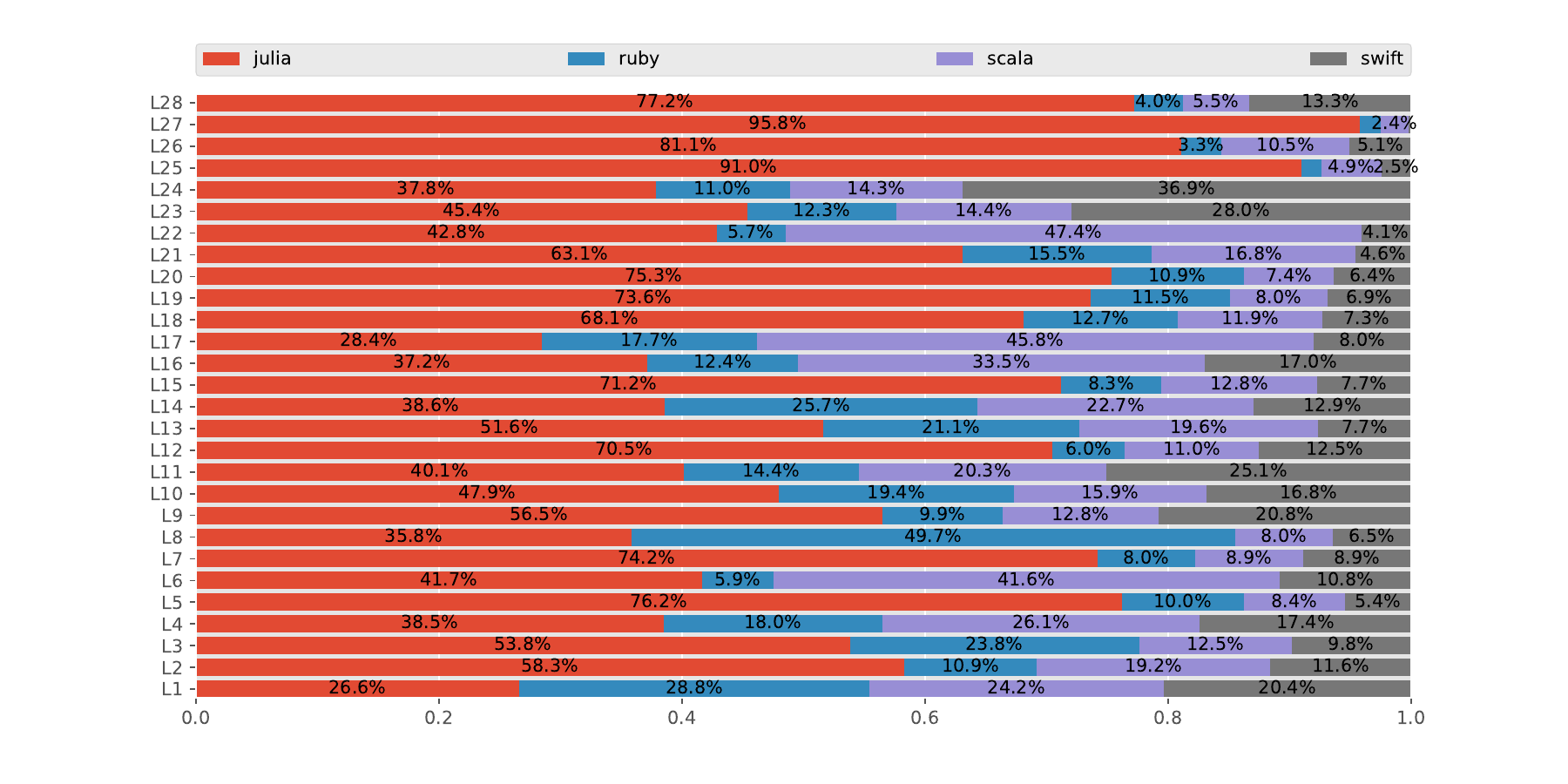}}}
    \caption{Attention to each target language observed on Qwen2.5-Coder 1.5B trained with AdapterFusion (top) and AdvFusion (bottom) for code translation to Julia. Attentions are collected over $18$ code translation samples with balanced source languages.}
    \label{fig:ct-bn-attn}
\end{figure}

\begin{figure}
    \centering
    \subfloat[\centering AdapterFusion+Compacter]{{\includegraphics[width=\textwidth]{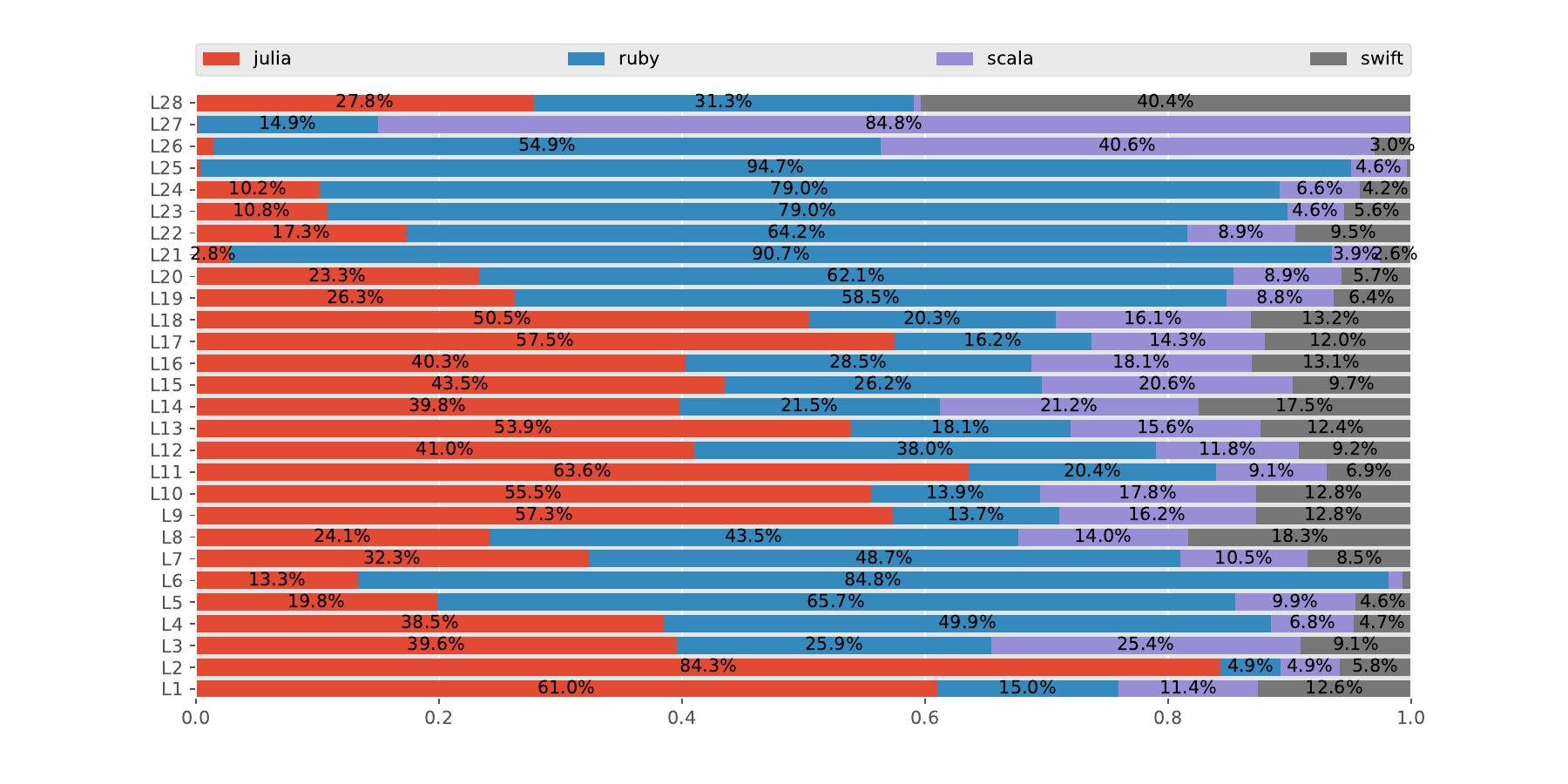}}}\\
    \subfloat[\centering AdvFusion+Compacter]{{\includegraphics[width=\textwidth]{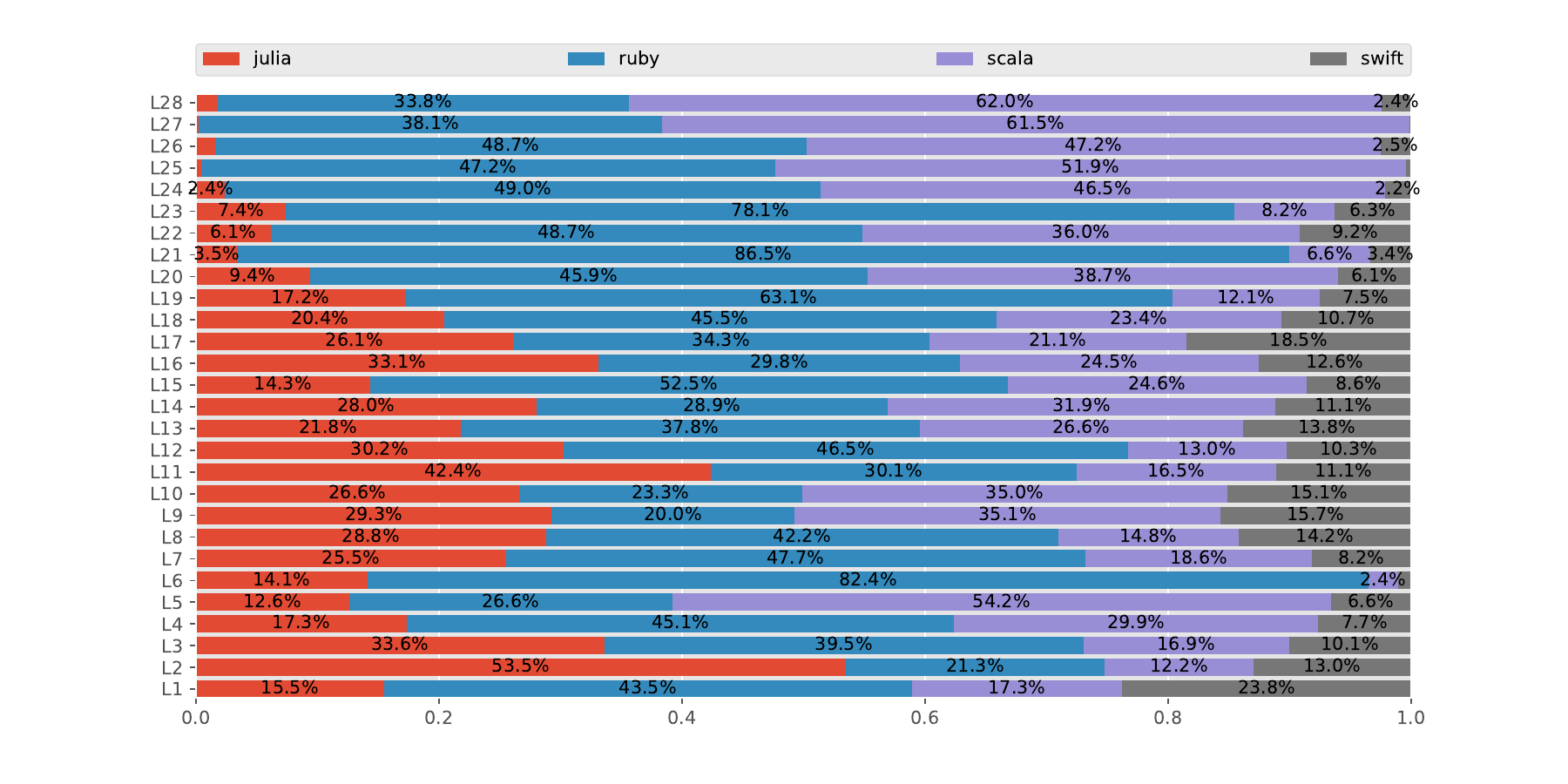}}}
    \caption{Attention to each target language observed on Qwen2.5-Coder 1.5B trained with AdapterFusion+Compacter (top) and AdvFusion+Compacter (bottom) for code translation to Julia. Attentions are collected over $18$ code translation samples with balanced source languages.}
    \label{fig:ct-comp-attn}
\end{figure}

Analyzing the attention distributions of the fusion layers reveals clear differences in how the models leverage programming language-specific adapters for code translation. Figure \ref{fig:ct-bn-attn} shows the attention distributions of the fusion layers across Qwen2.5-Coder 1.5B trained with AdapterFusion (top), AdvFusion (bottom), and Figure \ref{fig:ct-comp-attn} shows the attention distributions for AdapterFusion+Compacter (top) and AdvFusion+Compacter (bottom) when sampled on Julia code translation problems.
For AdvFusion, attention is highly concentrated on the Julia adapter in the upper layers, while the lower layers distribute attention more evenly across all programming languages. 
AdapterFusion spreads attention more evenly across programming languages, with a slight bias toward Julia in the middle layers and occasional peaks on other programming languages in the last layers. This is an interesting observation, as on small code-PLMs, the opposite effect is seen, where, compared to AdapterFusion, AdvFusion attends more to programming languages other than the target programming language.
Switching from bottleneck adapters to Compacter modules substantially changes attention dynamics. 
AdvFusion+Compacter initially focuses equally on all programming languages in the early and middle layers, shifting to Ruby and Scala in the top layers, where Julia receives little to no attention.
AdapterFusion+Compacter shows stronger attention to Julia in the early and middle layers, but, similar to  AdvFusion+Compacter, it transitions to Ruby and Scala in the upper layers.

These shifts in attention suggest that cross-language integration plays an important role in model effectiveness. In particular, AdvFusion+Compacter’s strong focus on both Julia and Ruby, two dynamic languages with similar syntactic structures, likely contributes to its superior functionality correctness. However, AdapterFusion+Compacter shows similar cross-language attention patterns but performs worse overall on Julia in code translation, indicating that attention allocation alone does not fully determine performance. Other factors, such as how the fusion architecture interacts with the adapters, also influence code translation effectiveness.

\subsection{Practical Insights}

While results from our previous study~\cite{saberi2025advfusion} established AdvFusion as the most effective fine-tuning strategy for smaller Code-LMs (pre-trained models) for code summarization and method name prediction, our results in this manuscript reveal that its advantages do not uniformly generalize to larger, autoregressive architectures and more challenging tasks, commit message generation, code generation, and code translation. The effectiveness of AdvFusion proves to be highly task, language, and model-dependent.

For CMG, AdvFusion remains competitive but no longer leads. It performs similarly to LoRA and TaskAdapter but is consistently outperformed by AdapterFusion, which achieves the best overall BLEU-4 and ROUGE-L scores across models. Replacing Bottleneck adapters with Compacter modules (AdvFusion+Compacter) further reduces performance, suggesting that adversarial fusion may not be well-suited for CMG. For practitioners, AdapterFusion emerges as the best overall method for CMG.

For code generation, AdvFusion surpasses AdapterFusion across all Code-LLMs, confirming that adversarial fusion helps integrate task-specific knowledge beneficial for code construction. However, it still falls behind TaskAdapter, which achieves the highest average BLEU-4 and ROUGE-L scores across models, followed closely by LoRA. Substituting Bottleneck adapters with Compacter modules produces mixed effects, helpful in mid-sized models but not consistently across the board. Thus, TaskAdapter remains the most effective and balanced method overall for code generation.

For code translation, AdvFusion’s limitations become more evident. Its performance declines significantly compared to LoRA and AdapterFusion, especially as the model size increases. However, this trend reverses when Compacter modules were introduced, since AdvFusion+Compacter substantially improves both functional correctness (Pass@1 and Pass@10) and textual similarity, often closing the gap with LoRA. This is not entirely unexpected, as Compacter itself proves highly capable on this task, performing on par with LoRA. While LoRA remains the best overall method for code translation, AdvFusion+Compacter offers a competitive alternative that achieves nearly comparable performance.

While AdvFusion showed superiority on smaller encoder-based models and simpler software engineering tasks~\cite{saberi2025advfusion}, its advantages are task-specific in larger decoder-only Code-LLMs. For practitioners, AdapterFusion and TaskAdapter remain high-performing choices for specific tasks such as CMG and code generation. AdvFusion+Compacter presents an intriguing new avenue, suggesting that adversarial mechanisms, when paired with compact parameterizations, can meaningfully enhance cross-lingual and structurally complex code-related tasks, such as code translation.

\section{Related Work}
\label{section:related-work}

We summarize related work for our study. We focus on three strands: (1) advances in PEFT and adapter composition, (2) up-to-date analyses of LLMs for code-driven software engineering tasks, and (3) recent task-specific studies for commit message generation, code generation and code translation.

\subsection{Parameter Efficient Fine Tuning Studies}

PEFT methods offer an alternative to fully fine-tuning language models and have been widely applied in NLP tasks\cite{houlsby2019parameter,pfeiffer2020mad,pfeiffer2020adapterfusion,hu2021lora}. Adapter-based fine-tuning is often shown to outperform full fine-tuning, particularly in zero-shot, cross-lingual, and low-resource scenarios \cite{low-res-related}. Meanwhile, several experiments have been conducted on various PEFT methods, such as Adapters, LoRA, and Prefix-Tuning, and they were evaluated on their performance, scalability, and knowledge transfer across more than 100 NLP tasks \cite{related-ding2023parameter}.

Research on PEFT approaches in software engineering is extensive \cite{goel2022cross,weyssow2023exploring,tse2023,liu2024mftcoder}. An empirical study on natural language to code transferability using adapters was conducted \cite{goel2022cross}. PEFT methods such as LoRA \cite{hu2021lora} and prompt tuning in code generation were also explored \cite{weyssow2023exploring}, with a focus on their advantages in large language models compared to small models. Prompt tuning's impact on CodeBERT and CodeT5 on code tasks such as defect prediction, summarization, and translation was investigated by Wang et al. \cite{tse2023}. They compared fully fine-tuned and prompt-tuned models, assessing accuracy and data efficiency. Other work proposed a multi-task fine-tuning framework using PEFT methods \cite{liu2024mftcoder}. The performance of PEFT approaches on Just-In-Time Defect Prediction (JIT-DP) and Commit Message Generation (CMG) is evaluated by Liu et al. \cite{liu2024delving}.

A recent work analyzed low-rank adaptation on the training dynamics and convergence regimes of LoRA,  explaining why LoRA typically finds useful low-rank solutions and when it may fail. This body of work provides theoretical grounding for using LoRA as a baseline in adapter comparisons \cite{kim2025lora_converge}. Other works introduced practical improvements to LoRA-style methods: progressive strategies (CoTo \cite{zhuang2025coto}, ProgLoRA \cite{yu2025prog_lora}) target adapter generalization and improved merging/pruning behaviour; LoRA-Gen \cite{xiao2025lora_gen} focuses on online generation of LoRA adapters to enable efficient specialization and deployment on edge devices. 

Newer PEFT methods, such as FLoE \cite{wang2025floe}, propose Fisher-guided, sparse layer selection to achieve improved adaptation efficiency and better layer-wise allocation of adapter capacity. There were also surveys and reviews of adapters that recommend best practices for PEFT design and evaluation across foundation models \cite{zhang2025peft_survey}. Other work related to adapter fusion and adapter-merging approaches examined progressive adapter schedules, merging/pruning robustness, and methods that improve adapter merging in multi-task or multilingual scenarios \cite{zhuang2025coto, yu2025prog_lora, xiao2025lora_gen}. 

These motivated our choice to evaluate multiple adapter families (LoRA, Compacter-like parameter-sharing adapters, and bottleneck-style adapters) and to test both standard fusion and adversarial fusion strategies on modern Code-LLMs.

\subsection{Language Models}

Recently, there has been multiple works on representing code using deep learning models for different applications such as code generation \cite{zeng2022extensive,zhou2022doccoder,fried2022incoder}, code summarization \cite{gu2022assemble,ahmed2022learning,nie2022impact}, program synthesis \cite{vaithilingam2022expectation,nijkamp2022conversational,ellis2021dreamcoder,austin2021program}, code search \cite{nadeem2022codedsi}, and bug fixing \cite{2022arXiv220700301R,2022arXiv220805446Z}.
A number of models were also released that are pre-trained on source code and/or code and comment with different objective functions, which are then fine-tuned on multiple downstream tasks \cite{wang2021codet5,feng2020codebert,guo2020graphcodebert} such as code summarization \cite{feng2020codebert,wang2021codet5,lu2021codexglue,ahmed2021multilingual}. Examples of these models include CodeT5 \cite{wang2021codet5}, CodeT5+ \cite{wang2023codet5+}, PLBART \cite{ahmad2021unified}, and CodeGPT. Each has versions fine-tuned for specific downstream tasks.

Recent advancements in code-LLMs have significantly advanced the capabilities of AI in software development.
Introduced by Meta, CodeLlama \cite{code-llama-roziere2023code} is a family of LLMs based on Llama2 \cite{touvron2023llama}, which is fine-tuned for code generation and understanding, and offers specialized versions for Python and instruction-based tasks. 
StarCoder \cite{li2023starcoder} is a 15.5B parameter model trained on 1 trillion tokens, featuring infilling capabilities and efficient large-batch inference enabled by multi-query attention.
CodeGemma \cite{team2024codegemma} is a collection of specialized open code models built on top of Gemma, capable of a variety of code and natural language generation tasks with excellent mathematical reasoning. 
Finally, the Qwen2.5-Coder \cite{hui2024qwen2} series includes models ranging from 0.5B to 32B parameters, built upon the Qwen2.5 architecture and pretrained on over 5.5 trillion tokens. This series of models demonstrates state-of-the-art performance across more than 10 benchmarks, including code generation, completion, reasoning, and repair.

\subsection{Task-specific Studies}

\paragraph{\textbf{Commit Message Generation}}
A recent work shows that LLMs can substantially help software engineering tasks but are highly sensitive to prompt design, data leakage, and evaluation choices \cite{liang2025swe}. Wu et al. \cite{wu2025cmg} found that in-context learning (ICL) with LLMs can produce competitive commit messages when prompts and datasets are carefully constructed and data leakage is controlled. Tsvetkov et al. \cite{tsvetkov2025realistic_eval} found that edit distance exhibits the highest correlation with their online edit-based metric, while BLEU \cite{papineni2002bleu} and METEOR \cite{banerjee-lavie-2005-meteor} correlate poorly. These insights directly informed our decision to include qualitative examples, to discuss metric limitations, and to both report lexical metrics and recommend embedding-based and user-centred evaluation in future work.

\paragraph{\textbf{Code Generation}}
Code-LMs have achieved advancements in code generation, enabling the generation of code from natural language descriptions \cite{jiang2024survey}. A variety of Code-LMs have recently been developed for code generation, including Codex \cite{chen2021evaluating}, CodeT5 \cite{wang2021codet5}, and CodeLlama \cite{roziere2023code}. These models enable generating code snippets based on provided natural language descriptions. Beyond this, researchers have explored approaches to improve generation quality and handle more complex coding tasks. For example, Zhang et al. \cite{zhang2023planning} proposed Planning-Guided Transformer Decoding (PG-TD), which uses lookahead search to guide the Transformer in generating higher-quality programs. Similarly, Jiang et al. \cite{jiang2024self} introduced a self-planning code generation approach, where the model first plans a sequence of solution steps and then generates code guided by these steps, improving correctness and robustness. Extending code generation to the repository level, Bairi et al. \cite{bairi2024codeplan} introduced CodePlan, which formulates repository-level coding as a planning problem and generates multi-step code edits while considering context from the entire codebase and previous changes. These works highlight the importance of code generation, which motivates us to focus on the code generation task. Instead of proposing new models, our work conducts empirical studies to explore how PEFT methods can improve the performance of Code-LLMs on code generation.

\paragraph{\textbf{Code Translation}}
Recent work on code translation has increasingly focused on developing new frameworks such as agentic systems, leveraging auxiliary signals like compiler feedback and runtime context, and scaling translation beyond single functions. A multi-agent LLM framework is introduced in \cite{yuan2024transagent}, where specialized agents collaborate to correct syntactic and semantic errors through alignment and execution feedback. Yin et al. \cite{yin2024rectifier} proposed a corrector model that repairs compilation, runtime, and functional errors in translated code, improving robustness across programming languages. Xin et al. \cite{xin2025enhancing} identified the challenges of long-sequence translation and introduced program state alignment to maintain functional equivalence over extended contexts. Jana et al. \cite{jana2023cotran} explored reinforcement learning with compiler and symbolic execution feedback to refine translation reliability, while Zhang et al. \cite{zhang2025scalable} demonstrated project-level translation pipelines that validate consistency across entire repositories.

In contrast to these architectural and validation-oriented approaches, our work systematically evaluates a range of PEFT methods for code translation, with a particular focus on functional correctness. We analyzed how adapter-based strategies, including AdvFusion, a recent fusion-based PEFT technique, affect translation correctness and textual fidelity, offering complementary insights into fine-tuning efficiency rather than proposing new translation architectures.

\section{Threats to Validity}
\label{section:threats}
\paragraph{\textbf{Internal Threats}}
An internal threat can arise from the use of more efficient data types and quantization configurations. We employ \texttt{bfloat16} precision and 4-bit quantization to enable efficient fine-tuning while maintaining practical feasibility across all experiments. To mitigate this threat, we ensure consistency by training all setups under the same configuration. Additionally, low-bit quantization has become mainstream in modern large model fine-tuning pipelines, which helps reduce the risk of confounding effects. Nevertheless, our results may not fully generalize to scenarios using full-precision or unquantized models.
Hyperparameter and implementation sensitivity may constitute another internal threat. We address this by adhering to the default hyperparameters recommended by the original authors of each PEFT method, which have been extensively validated in prior work. Furthermore, we rely on widely adopted implementations from established libraries to ensure reproducibility and correctness.

\paragraph{\textbf{External Threats}}
External threats relate to the generalizability of our findings beyond the studied tasks. Our experiments cover three representative and complex software engineering tasks that span diverse programming language modalities. This helps improve the robustness of our conclusions. However, despite this diversity, the results may not directly generalize to other software engineering tasks or modalities.

\paragraph{\textbf{Construct Threats}}
A potential construct threat involves the generalization of results to other model families. We mitigate this by including a diverse set of recent models that vary in architecture type and parameter budget, thereby ensuring a broader coverage of model characteristics.
Another construct threat concerns the possible misrepresentation of performance due to metric selection. For commit message generation, where the primary goal is fidelity to the ground truth, we rely on textual similarity metrics such as BLEU-4 and ROUGE-L. In contrast, for code translation, where multiple correct implementations may exist, we complement textual metrics with functionality-based measures such as Pass@1 and Pass@10. Finally, for code generation, although functionality-based evaluation was performed, all functional correctness scores were zero, so we refrain from reporting them to avoid repetition.

\section{Conclusion and Future Works}
\label{section:conclusion}

We extend the prior work of AdvFusion \cite{saberi2025advfusion} by exploring the performance of AdvFusion on Code-LLMs by focusing on three new tasks, including code generation, commit message generation and code translation. We also compared AdvFusion with several popular PEFT methods such as LoRA, Compacter, and AdapterFusion. The results showed that different tasks exhibit different characteristics.
In code generation, the results show that AdvFusion achieves better performance than AdapterFusion. However, other PEFT methods, such as LoRA, Compacter and TaskAdapter, AdvFusion generally achieves lower performance. Replacing Bottleneck adapters with Compacter does not lead to improvements for AdvFusion. Overall, while AdvFusion provides improvements over AdapterFusion, LoRA, and TaskAdapter remain more robust and high-performing baselines for code generation. 

In commit message generation, AdapterFusion frequently matches or surpasses AdvFusion on recent code-LLMs, particularly with certain adapter designs. LoRA and TaskAdapter remain robust, high-performing baselines, while Compacter is competitive in selected settings. We also find that adapters trained directly on commit-message pairs outperform adapters trained only on code for the CMG objective, underscoring the importance of task-aligned data and adapter architecture choices.  
In code translation, AdapterFusion generally outperforms AdvFusion, and replacing bottleneck adapters with Compacter can improve AdvFusion’s performance in some cases. Consistent with code generation and commit message generation, LoRA demonstrates competitive performance.

For future work, we recommend several complementary directions to stre-ngthen and broaden these findings: (1) conduct systematic robustness studies of low-bit quantization and optimizer variants to address instability observed on Code-LLMs and to better characterize 4-bit training dynamics for AdvFusion \cite{kim2025lora_converge, bitsandbytes2023}; (2) evaluate recent PEFT methods (progressive activation schedules such as CoTo, FLoE, and layer-aware adapter placement) in the AdapterFusion and AdvFusion pipelines to improve transfer and stability \cite{zhuang2025coto, wang2025floe}; (3) and complement automatic metrics with semantic and human-centered evaluations to ensure generated commit messages are not only lexically close but practically useful to developers \cite{wu2025cmg, tsvetkov2025realistic_eval}. These directions will help translate our empirical insights into robust, deployable solutions.

\section*{Declarations}

\subsection*{Funding}
This research is supported by grants from the Natural Sciences and Engineering Research Council of Canada, RGPIN-2019-05175 and ALLRP 590428-23.

\subsection*{Data Availability Statement}
We include all scripts and tooling used to obtain the results in our GitHub repository\footnote{\url{https://github.com/Amirresm/advfusion-cllm}}.

\subsection*{Conflict of Interest}
The authors declare that they have no conflict of interest.

\balance
\bibliographystyle{unsrt}
\bibliography{References}

\end{document}